\documentclass[a4paper,11pt]{article}
\pdfoutput=1 

\usepackage{jcappub}
\usepackage{hyperref}
\usepackage[noabbrev]{cleveref}
\usepackage{mathtools}
\graphicspath{ {plots/} }
\usepackage{subcaption}
\usepackage{graphicx}
\usepackage{comment}
\usepackage{orcidlink}

\newcommand{\DS}{\mathrm{DS}}
\newcommand{\hMpc}{h^{-1}\mathrm{Mpc}}
\newcommand{\Mpch}{h\mathrm{Mpc}^{-1}}
\newcommand{\hGpc}{h^{-1}\mathrm{Gpc}}

\newcommand{\xiRDS}{\xi_{R}^{\DS}}

\newcommand{\xiRRDS}[2]{\xi_{R_{#1}, R_{#2}}^{\DS}}
\newcommand\vc[1]{\mathbf{#1}}
\newcommand\vx{\vc{x}}
\newcommand\vy{\vc{y}}
\newcommand\vr{\vc{r}}
\newcommand\vs{\vc{s}}
\newcommand\vk{\vc{k}}
\newcommand{\dtilde}{\tilde{\delta}_{R}^{\DS}}

\newcommand{\xdeltaR}{\tilde{\delta}_{R}}
\newcommand{\prob}{\mathcal{P}}
\newcommand{\poisson}{\mathrm{Pois}}

\newcommand{\nbar}{\bar{n}}
\newcommand{\eff}{\mathrm{eff}}
\newcommand{\xiYR}{\xi_{Y_R}}
\newcommand{\sigmaYR}{\sigma_{Y_R}}
\newcommand{\rhoYR}{\rho_{Y_R}}
\newcommand{\xiRgDS}{\xi_{R, g}^{\DS}}

\title{\boldmath A theoretical approach to density-split clustering}

\author[a, 1]{Mathilde Pinon\orcidlink{0009-0009-3228-7126},\note{Corresponding author.}}
\author[a]{Arnaud de Mattia\orcidlink{0000-0003-0920-2947},}
\author[a]{Étienne Burtin\orcidlink{0000-0002-0302-5145},}
\author[a]{Vanina Ruhlmann-Kleider\orcidlink{0009-0000-6063-6121},}
\author[b]{Sandrine Codis\orcidlink{0000-0002-3037-0032},}
\author[c]{Enrique Paillas\orcidlink{0000-0002-4637-2868},}
\author[d,e,f]{Carolina Cuesta-Lazaro\orcidlink{0000-0002-6069-2999}}

\affiliation[a]{Universit\'e Paris-Saclay, CEA, IRFU,  F-91191 Gif-sur-Yvette, France}
\affiliation[b]{Universit\'e Paris-Saclay, Université Paris Cité, CEA, CNRS, AIM, 91191, Gif-sur-Yvette, France}
\affiliation[c]{Department of Astronomy/Steward Observatory, University of Arizona, 933 North Cherry
Avenue, Tucson, AZ 85721-0065, USA}
\affiliation[d]{The NSF AI Institute for Artificial Intelligence and Fundamental Interactions, Cambridge,
MA 02139, USA}
\affiliation[e]{Department of Physics, Massachusetts Institute of Technology, Cambridge, MA 02139, USA}
\affiliation[f]{Center for Astrophysics --- Harvard \& Smithsonian, 60 Garden Street, MS-16, Cambridge, MA 02138, USA}

\emailAdd{mathilde.pinon@cea.fr}

\abstract{We present an analytical model for density-split correlation functions, that probe galaxy clustering in different density environments. Specifically, we focus on the cross-correlation between density-split regions and the tracer density field. We show that these correlation functions can be expressed in terms of the two-point probability density function (PDF) of the density field. We derive analytical predictions using three levels of approximation for the two-point PDF: a bivariate Gaussian distribution, a bivariate shifted log-normal distribution, and a prediction based on the Large Deviation Theory (LDT) framework. For count-in-cell densities, obtained through spherical top-hat smoothing, one can leverage spherical collapse dynamics and LDT to predict the density two-point PDF in the large-separation regime relative to the smoothing radius. We validate our model against dark matter N-body simulations in real space, incorporating Poisson shot noise and galaxy bias. Our results show that the LDT prediction outperforms the log-normal approximation, and agrees with simulations on large scales within the cosmic variance of a typical DESI DR1 sample, despite relying on only one degree of freedom.}

\begin{document}
\maketitle
\flushbottom

\section{Introduction}
\label{sec:intro}

Galaxy clustering is a powerful tool for probing the large-scale structure of the Universe, providing critical insights into cosmological parameters driving its evolution. During the Universe's early stages, a rapid expansion phase known as inflation generated nearly Gaussian fluctuations in the matter density field, as evidenced by observations of the cosmic microwave background. However, as the Universe evolved, gravitational collapse introduced non-linear effects, causing the matter density fluctuations to become increasingly non-Gaussian~\cite{bernardeau_large-scale_2002}. 

Current galaxy clustering analyses predominantly rely on two-point statistics, such as the two-point correlation function or its Fourier counterpart, the power spectrum, to compress the information contained in the galaxy density field. While efficient, these methods are limited to capturing the variance of the density field across a range of scales, thereby under-utilizing the wealth of information encoded in modern surveys. As large-scale structure surveys expand, with for instance the Dark Energy Spectroscopic Instrument (DESI)~\cite{desi_collaboration_desi_2016}, Euclid~\cite{laureijs_euclid_2011}, or the upcoming Vera Rubin LSST~\cite{ivezic_lsst_2019}, it becomes crucial to develop more efficient methods for extracting information.

Recent years have seen a growing interest in exploring higher-order statistics, like the three-point correlation function~\cite{sugiyama_new_2023}, bispectrum~\cite{philcox_boss_2022} or higher-order correlation functions~\cite{philcox_first_2021}. Alternative summary statistics, such as the marked power spectrum~\cite{massara_using_2021, ebina_analytically_2024}, Minkowski functionals~\cite{liu_probing_2023, jiang_minkowski_2024}, $k$-nearest neighbors~\cite{banerjee_nearest_2021}, voids~\cite{nadathur_beyond_2019, hawken_constraints_2020}, or wavelet scattering transforms~\cite{valogiannis_towards_2022, valogiannis_precise_2023}, are also gaining traction as efficient and informative probes of galaxy clustering. Another promising approach is field-level inference~\cite{lavaux_systematic-free_2019, andrews_bayesian_2023}, which models the entire 3D density field directly rather than relying on summary statistics. Comparative studies of these methods, such as that by the "Beyond-2pt" collaboration~\cite{collaboration_parameter-masked_2024}, highlight their growing relevance.

Among these alternative methods, \textit{density-split clustering} has emerged as a particularly promising method~\cite{paillas_redshift-space_2021, paillas_constraining_2023, paillas_cosmological_2023, cuesta-lazaro_sunbird_2024, morawetz_constraining_2024}. By measuring galaxy clustering within distinct local density environments, it provides a physically interpretable probe for exploring higher-order information in the density field. A recent study~\cite{paillas_cosmological_2023} has successfully applied density-split clustering for cosmological inference on the BOSS CMASS galaxy sample, demonstrating its strong constraining power on cosmological parameters. Density-split statistics for weak lensing, which measure cosmic shear around split foreground environments, have also been applied to data~\cite{gruen_density_2018, burger_kids-1000_2023} and modeled analytically~\cite{friedrich_density_2018}.

Advances in simulations -- both in resolution and volume -- and machine learning techniques have enabled accurate modeling of many summary statistics using simulation-based emulators. These emulators cover a wide range of scales and may incorporate observational systematics. In particular, previous density-split clustering analyses have relied on a simulation-based model~\cite{cuesta-lazaro_sunbird_2024}, as no analytical model exists yet. However, simulation-based methods require great confidence in the simulations, and often rely on numerous parameters that may be difficult to link to the underlying physical processes. Besides, generating sufficiently large simulations to match the volume of ongoing surveys like DESI presents a significant challenge.

In this work, we present several approaches for deriving analytical predictions of density-split correlation functions, also referred to simply as \textit{density-splits} for brevity. The most straightforward method assumes a Gaussian underlying density field, and we derive the corresponding density-split correlation functions under this approximation. Furthermore, leveraging the fact that the matter density field is approximately log-normal, we develop an analytical expression for density-splits based on this log-normal assumption.

An alternative lies in count-in-cell statistics -- which measure the distribution of galaxy (or other tracer) counts within spherical cells of a given radius. Notably, their one-point probability distribution function (PDF), can be modeled with high accuracy using the Large Deviation Theory (LDT) formalism~\cite{bernardeau_statistics_2014, bernardeau_joint_2015, bernardeau_large_2016, uhlemann_back_2016, uhlemann_beyond_2017, codis_encircling_2016, codis_large-scale_2016}. Unlike standard perturbation theory, which assumes small density contrasts everywhere, LDT requires only the density variance to be small. This framework predicts that the PDF has an exponential decay governed by a so-called rate function, which can be predicted under some symmetric configurations, for instance with spherical collapse dynamics.

Building on the LDT, we derive an analytical model for density-split correlation functions. In particular, we use the extension introduced in~\cite{uhlemann_back_2016}, that applies the LDT to the logarithmic transform of the count-in-cell density, thereby extending LDT's applicability to variances approaching 1. We connect the density-split correlation functions to the one-point and two-point PDF of the density contrast, as well as its bias function in the large separation regime, and use the large-separation limit LDT prediction for the bias function developed in~\cite{codis_large-scale_2016}. To validate our approach, we compare our model against measurements from AbacusSummit N-body simulations~\cite{garrison_abacus_2021, maksimova_abacussummit_2021}.

This paper is organized as follows. Section~\ref{sec:density-split_clustering} defines density-split correlation functions. Section~\ref{sec:gaussian_density} introduces an analytical Gaussian model. Section~\ref{sec:measurements_abacus} describes the simulations and estimators used for validation. Finally, section~\ref{sec:model} presents analytical descriptions for the matter field based on log-normal assumption and LDT, and section~\ref{sec:biased_tracers} extends this framework to biased tracers of matter. Conclusions are summarized in section~\ref{sec:conclusion}.

All the code used in this work is publicly available.\footnote{\url{https://github.com/mathildepinon/densitysplit}.}

\section{Density-split clustering}\label{sec:density-split_clustering}
Density-split correlation functions are analogous to the standard two-point correlation function, but are defined with respect to a local density environment. We call \textit{density-split} (DS) a region defined by a given range of density. Different types of correlation functions may be of interest, for instance the auto-correlation of random positions within the DS, or the auto-correlation of the particles within each DS. Previous works on density-split clustering \cite{paillas_redshift-space_2021, paillas_constraining_2023, paillas_cosmological_2023} used the cross-correlation of random points from each DS with the density field, as well as the auto-correlation functions of the $\DS$, which \cite{paillas_redshift-space_2021} showed to add useful cosmological information when computing density-splits in redshift space. In this work, we focus on the cross-correlation of random points from each DS with the density field, but a similar approach can easily be used to obtain the auto-correlation functions of the $\DS$.

In practice, we measure the density from a particle catalog using a kernel function $K_{R}$ characterized by a smoothing scale $R$
\begin{equation}
\label{eq:smoothed_density_contrast}
\delta_{R}(\vr) = \int d^3\vx \, K_{R}(\vr, \vx) \delta(\vx),
\end{equation}
where $\delta$ denotes the density contrast $\delta(\vx) = \frac{n(\vx) - \Bar{n}}{\Bar{n}}$, $n(\vx)$ being the number density of particles and $\Bar{n}$ the average number density. The cross-correlation function of random points in the density-split DS with the whole field of particles, $\xiRRDS{1}{2}$ is 
formally defined as the conditional expectation of the particle density at $\vr + \vs$ given the particle density at $\vr$:
\begin{equation}
\xiRRDS{1}{2}(\vs)
= \langle \delta_{R_2}(\vr + \vs) | \delta_{R_1}(\vr) \in \DS \rangle
\label{eq:cross_corr_randoms_halos_expectation}
\end{equation}
i.e.
\begin{equation}
1 + \xiRRDS{1}{2}(\vs) =
\frac{1}{|\DS|} \int_{\mathrm{DS}} d\delta_{R_1}(\vr) 
\int_{-1}^{+\infty} d\delta_{R_2}(\vr + \vs)
\left( 1 + \delta_{R_2}(\vr + \vs) \right)
\prob(\delta_{R_1}(\vr), \delta_{R_2}(\vr + \vs))
\label{eq:cross_corr_randoms_halos}
\end{equation}
where
\begin{equation}
    |\DS| = \int_{\delta_R \in \mathrm{DS}} d\delta_R \,\prob(\delta_R).
\label{eq:DS_def}
\end{equation}

Thus, one of the key components of the density-split correlation function $\xiRRDS{1}{2}(\vs)$ is the joint PDF of $\delta_{R_1}(\vr)$ and $\delta_{R_2}(\vr + \vs)$. Note that if we were considering the auto-correlation of each DS, we would drop the $\delta_{R_2}(\vr + \vs)$ term and integrate over $\delta_{R_2}$ only in the DS region, and if we were considering the cross-correlation of the particles of each DS with all the particles, there would be an additional $(1 + \delta_{R_1}(\vr))$ factor inside the integral. Also note that here $R_1$ and $R_2$ can take any values. One could additionally consider the case where $R_2 \xrightarrow{} 0$, replacing $\delta_{R_2}$ by $\delta$. In this work, for simplicity, we consider $R_1 = R_2 = R$ (although all the formulae can be easily extended to the case where $R_1 \neq R_2$). Therefore, in the following we will write:
\begin{equation}
    \xiRDS(\vs) = \xiRRDS{}{}(\vs).
\end{equation}

\section{Gaussian density}\label{sec:gaussian_density}
In this section, to get an intuition of the information encoded in the density-split correlation function,  we assume a Gaussian distribution for the density contrast, i.e. that
\begin{equation}
    \prob(\delta_{R}(\vr), \delta_{R}(\vr + \vs)) \sim \mathcal{G}(0, \Sigma_{R}(\vs)),
\end{equation}
with
\begin{equation}
\label{eq:smoothed_cov_matrix}
\Sigma_{R}(\vs) = 
\begin{pmatrix}
    \sigma_{R}^2 & \xi_{R}(\vs)\\
    \xi_{R}(\vs) & \sigma_{R}^2
\end{pmatrix}  ,
\end{equation}
where $\sigma_{R}^2 = \langle \delta_{R}(0)^2 \rangle$ is the variance of the smoothed density contrast and $\xi_{R}(\vs) = \langle \delta_{R}(\vr)\delta_{R}(\vr+\vs) \rangle = \int d^3\vx \int d^3 \vy\, K_{R}(\vs, \vx)K_{R}(\vx, \vy)\xi(\vy)$ is the smoothed two-point correlation function. Thus we can derive an analytical expression for~\cref{eq:DS_def,eq:cross_corr_randoms_halos} 
\begin{equation}
\label{eq:n_R_DS_avg_computed}
|\DS| = \frac{1}{2} 
\left[ \mathrm{erf} \left(\frac{\delta}{\sqrt{2}\sigma_{R}}\right) 
\right]_{\delta_{1}}^{\delta_{2}}
\end{equation}
and
\begin{equation}
\label{eq:cross_corr_randoms_halos_computed}
|\DS| \left( 1 + \xiRRDS{1}{2}(\vs) \right) =
\frac{1}{2} 
\left[ \mathrm{erf} \left(\frac{\delta}{\sqrt{2}\sigma_{R}}\right) 
\right]_{\delta_{1}}^{\delta_{2}}
- \frac{\xi_{R}(s)}{\sqrt{2\pi} \sigma_{R}} \left[ \exp \left( - \frac{\delta^2}{2\sigma_{R}^2} \right)  
 \right]_{\delta_{1}}^{\delta_{2}}
\end{equation}
where $\DS$ is the interval $[\delta_1, \delta_2]$. Combining equations~\eqref{eq:n_R_DS_avg_computed} and \eqref{eq:cross_corr_randoms_halos_computed}, we get
\begin{equation}
\label{eq:xi_R_DS_computed}
\xiRDS(s)
= - \sqrt{\frac{2}{\pi}} \frac{\xi_{R}(s)}{\sigma_{R}}
\frac{
\left[ \exp \left( - \frac{\delta^2}{2\sigma_{R}^2} \right)
\right]_{\delta_{1}}^{\delta_{2}}
}{
\left[ \mathrm{erf} \left(\frac{\delta}{\sqrt{2}\sigma_{R}}\right) 
\right]_{\delta_{1}}^{\delta_{2}}
}.
\end{equation}
In the above equation we recognize the average smoothed density contrast in DS, $\dtilde$:
\begin{equation}
\label{eq:delta_R_tilde}
\dtilde
= \frac{\int_{\mathrm{DS}} d\delta 
\delta \exp \left( - \frac{\delta^2}{2\sigma_{R}^2} \right)}{\int_{\mathrm{DS}} d\delta \exp \left( - \frac{\delta^2}{2\sigma_{R}^2} \right)}
= - \sigma_{R} \sqrt{\frac{2}{\pi}} \frac{
\left[ \exp \left( - \frac{\delta^2}{2\sigma_{R}^2} \right)
\right]_{\delta_{1}}^{\delta_{2}}
}{
\left[ \mathrm{erf} \left(\frac{\delta}{\sqrt{2}\sigma_{R}}\right) 
\right]_{\delta_{1}}^{\delta_{2}}
}
\end{equation}
such that $\xiRDS(\vs)$ can be written as a function of the smoothed two-point correlation function $\xi_{R}(\vs)$, the variance of the smoothed density contrast $\sigma_{R}^2$ and $\dtilde$:
\begin{equation}
\label{eq:density_splits_gaussian_model}
\xiRDS(\vs) = \frac{\dtilde}{\sigma_{R}^{2}} \xi_{R}(\vs). 
\end{equation}
Equation~\eqref{eq:density_splits_gaussian_model} shows that at first order, the density-split correlation function differs from the smoothed two-point correlation function only by a linear bias factor proportional to the average density in the given region. It is analogous to the result derived in~\cite{kaiser_spatial_1984} for clusters (often referred to as Kaiser bias in the literature), which we extend here for density-split correlation functions.

Figure~\ref{fig:density_splits_gaussian} shows that this Gaussian prediction does not hold in the general case. Even when using $\dtilde$ and $\sigma_{R}^2$ measurements from dark matter simulations (see below) in the above equation~\eqref{eq:density_splits_gaussian_model}, the model fails to reproduce the simulations at scales below $120 \; \hMpc$. 

Note that here, contrary to section~\ref{sec:model}, we did not take shot noise into account. The aim was to derive a simple analytical expression in the purely Gaussian case. However, in figure~\ref{fig:density_splits_gaussian}, shot noise is naturally included in the quantities $\dtilde$, $\sigma_{R}^2$ and $\xi_{R}(\vs)$, since they are measured from the simulations.

\section{Measurements on AbacusSummit simulations}\label{sec:measurements_abacus}

\subsection{AbacusSummit simulations}

To test our different models, we compare them against dark matter AbacusSummit N-body simulations at baseline cosmology, which is Planck $\Lambda$CDM cosmology~\cite{planck_collaboration_planck_2020}. We use 25 realizations of periodic cubic box simulations with $2 \; \hGpc$ side length at redshift $z = 0.8$. Dark matter particles are downsampled to $\nbar = 3.4 \times 10^{-3}  \; (\Mpch)^3$. To get an estimation of typical error bars corresponding to current surveys like DESI, we additionally compute the density and density-splits of the simulations down-sampled to $\nbar = 5 \times 10^{-4} \; (\Mpch)^3$. This is the approximate number density of the $\texttt{LRG3 + ELG1}$ sample in DESI first data release (DR1) at redshift $z \sim 0.8$ (with $dz = 0.2$)~\cite{desi_collaboration_desi_2024, collaboration_desi_2024}. Given that the volume of each AbacusSummit simulation is $V = 8 \; (\hGpc)^3$, the effective volume of the down-sampled simulations is:
\begin{equation}
    V_{\eff} = \left( \frac{\nbar P_{0}(k = 0.14 \; \Mpch)}{1 + \nbar P_{0}(k = 0.14 \; \Mpch)} \right)^2 V \simeq 2.5 \; (\hGpc)^3
\end{equation}
where $P_{0}(k = 0.14 \; \Mpch) \simeq 2400 \; (\hMpc)^3$ is the power spectrum monopole at $k = 0.14 \; \Mpch$ evaluated on the simulations. This effective volume roughly matches the approximate $2 \; (\hGpc)^3$ volume of the $\texttt{LRG3 + ELG1}$ sample.

\subsection{Count-in-cell density estimator}\label{sec:cic_density_estimator}

In principle, the kernel $K_R$ in equation~\eqref{eq:smoothed_density_contrast} could be any smoothing kernel. However, the LDT framework that we will use in section~\ref{sec:ldt} is based on spherical collapse predictions, which have been shown to perform well in the case of a spherical top-hat kernel. Hence, in the following, we use a spherical top-hat kernel with radius $R$
\begin{equation}
    K_{R}(\vr, \vx) = 
    \begin{cases}
        \frac{3}{4 \pi R^3}  & \mathrm{if} \; \lVert \vr - \vx \rVert \leq R \\
        0  & \rm otherwise
    \end{cases}
\end{equation}
or in Fourier space:
\begin{equation}
    K_{R}(k) = 3 \left( \frac{\sin(Rk)}{(Rk)^3} - \frac{\cos(Rk)}{(Rk)^2} \right),
\end{equation}
where $k = \lVert \vk \rVert$. Thus, in what follows, $\delta_R$ refers to the \textit{count-in-cell} density. 

A possible method to compute the count-in-cell density is to interpolate the particle positions on a 3D mesh, then perform a 3D Fast Fourier Transform (FFT), apply the top-hat smoothing in Fourier space, and then reverse FFT back to configuration space. However in that case, the resulting smoothing kernel also depends on the interpolation scheme used to assign the particles positions to the mesh, and may deviate from the spherical symmetry assumption. In order to eliminate errors due to an inexact smoothing kernel, we apply the top-hat smoothing directly in configuration space. In practice, we define a mesh with cell size $5 \; \hMpc$, and compute the number of particles falling within a radius $R$ of each of the mesh nodes, which yields count-in-cell measurements for $64 \times 10^6$ spherical cells. Since the $R$ values that we use are larger than $5 \; \hMpc$, the cells are partially overlapping.

To get an estimator for the density contrast, we normalize our count-in-cell measurements by the average number density of the 25 simulations and retrieve 1. In order to have the same binning in $\delta_R$ for all simulations, we normalize each simulation with the overall average number density of the 25 simulations (rather that its own average number density). Thus, we get a probability estimation for discrete values of the density contrast. In all the figures showing the density PDF, the blue circles represent measurements from simulations and correspond to $N=1, 2, 3...$ particles per sphere of radius $R$, form left to right. The spacing between the points is determined by the average number density of the simulations and the smoothing radius under consideration.

Figure~\ref{fig:one-point_pdf} shows the average count-in-cell density PDF in the 25 AbacusSummit simulations with $\nbar = 3.4 \times 10^{-3}  \; (\Mpch)^3$ (blue circles). The blue area in the bottom panel shows the standard deviation of the 25 simulations, which is around 1\textperthousand\ of the PDF in the region $\delta_R \simeq 0$. The measured variance of the smoothed density contrast is $\sigma_R^2 = 0.30$, or $0.23$ when retrieving the contribution from a Poisson shotnoise. Figure~\ref{fig:one-point_pdf_lownbar} shows the density PDF measured in the 25 simulations with $\nbar = 5 \times 10^{-4}  \; (\Mpch)^3$, with two different smoothing radii: $R = 10 \; \hMpc$ and $R = 25 \; \hMpc$. Note that, as expected, the larger the smoothing scale, the smaller the variance, and the closer to Gaussian the PDF is.

\subsection{Density-split correlation function estimator}

The density-split correlation function is computed using the standard Landy-Szalay estimator:
\begin{equation}
    \widehat{\xiRDS}(\vs) = \frac{\rm R^{DS} D(\vs) - R^{DS} R(\vs) - R D(\vs) + RR(\vs)}{\rm RR(\vs)}
\end{equation}
where $\rm XY (\vs)$ denotes pair counts. Here R represents the positions of the mesh nodes, $\rm R^{DS}$ are the positions of the mesh that fall within the density-split region DS, and $\rm D$ are the positions of the mesh weighted by the measured smoothed density contrast at these positions. As we work in real space, the density is isotropic so we only consider the monopole of the correlation function. We measure the correlation function monopole for 50 linearly spaced bins in the range $s \in [0, 150] \; \hMpc$. The correlation functions (either $\xi_R$ or $\xiRDS$) are computed with the Python package \texttt{pycorr}\footnote{\url{https://github.com/cosmodesi/pycorr}}, which is based on \texttt{Corrfunc}~\cite{sinha_corrfunc_2019, sinha_corrfunc_2020}.

\begin{figure}
\centering 
\includegraphics[scale=0.8]{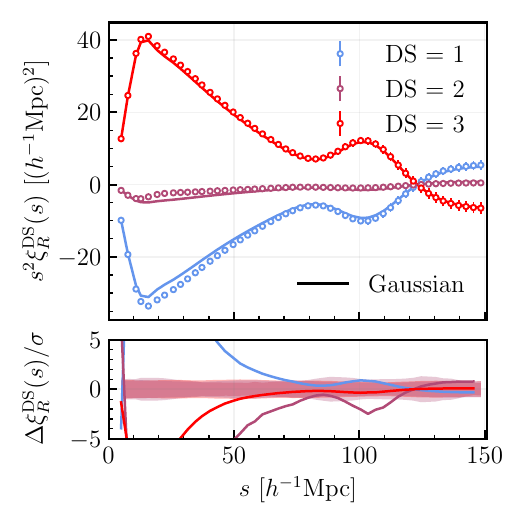}
\caption{Average density-split correlation functions of the 25 AbacusSummit simulations with $\nbar = 0.0034 \; (\Mpch)^3$ and $R = 10 \; \hMpc$ (circles), compared with the Gaussian approximation from equation~\eqref{eq:density_splits_gaussian_model} (solid lines). Each of the three quantities $\dtilde$, $\xi_R$ and $\sigma_R$ that appear in equation~\eqref{eq:density_splits_gaussian_model} is averaged over the 25 simulations. Bottom panel shows the difference between the Gaussian approximation and the simulations, divided by the standard deviation of the 25 mocks. The shaded areas represent the standard deviation from the low density simulations ($\nbar = 5 \times 10^{-4} \; (\Mpch)^3$) divided by the standard deviation from the high density simulations ($\nbar = 0.0034 \; (\Mpch)^3$), for the three density-splits. It shows approximately the 1$\sigma$ errors we would expect from a sample with a number density and effective volume similar to that of DESI DR1 \texttt{LRG3+ELG1} bin.}
\label{fig:density_splits_gaussian}
\end{figure}

The density-splits can be any density intervals, but here we choose to define them as quantiles of the density PDF, in line with what was done in previous density-split analyses~\cite{paillas_cosmological_2023}. To ensure that the density intervals are fixed for all simulations and for the models, we compute the quantiles from the log-normal approximation of the density PDF (see section~\ref{sec:log-normal}) of the first simulation. For instance, with 3 quantiles, the $\DS$ boundaries are approximately: [-1, -0.29, 0.11, $+\infty$]; and with 5 quantiles: [-1, -0.44, -0.22, 0.02, 0.37, $+\infty$]. Figure~\ref{fig:density_splits_gaussian} shows the density-split correlation functions with 3 quantiles computed on the simulations, together with the Gaussian approximation from equation~\eqref{eq:density_splits_gaussian_model}.

\subsection{Bias function estimator}

Let us introduce the \textit{bias function} $b(\delta_R, \vs)$, that characterizes the expansion of the two-point PDF $\prob(\delta_R(\vr), \delta_R(\vr + \vs))$ around $\prob(\delta_R(\vr))\prob(\delta_R(\vr + \vs))$ in the limit of large separation through~\cite{codis_large-scale_2016, uhlemann_beyond_2017, uhlemann_it_2023}:
\begin{equation}
    \prob(\delta_R, \delta_R^{\prime}) = \prob(\delta_R) \prob(\delta_R^{\prime}) \left[ 1 + \xi_R(\vs) b(\delta_R, \vs) b(\delta_R^{\prime}, \vs) \right]
\label{eq:bias_peak_background}
\end{equation}
where we noted $\delta_R = \delta_R(\vr)$, $\delta_R^{\prime} = \delta_R(\vr + \vs)$ for better readability. We will see in section~\ref{sec:ldt} that in this large-separation limit, $s \gg R$, the bias function becomes independent of separation and reduces to $b(\delta_R)$, such that we can interpret equation~\eqref{eq:bias_peak_background} as a \textit{peak-background split} analog for count-in-cells.
From equation~\eqref{eq:bias_peak_background}, and by imposing a normalization condition on the bias function to ensure the defintion of $\xi_R$ (equation~\eqref{eq:bias_normalization}), we can derive:
\begin{equation}
\begin{split}
    b(\xdeltaR, \vs) \xi_R (\vs) 
    & = \langle \delta_R(\vr + \vs) | \delta_R(\vr) = \xdeltaR \rangle \\
    & = \frac{\int_{-1}^{+\infty} d\delta_R(\vr + \vs) \delta_R(\vr + \vs) \prob(\delta_R(\vr)  = \xdeltaR, \delta_R (\vr + \vs))}{\prob(\delta_R(\vr) = \xdeltaR)}.
\end{split}
\label{eq:bias_function}
\end{equation}

Then we can write the density-split correlation function (as defined by equation~\eqref{eq:cross_corr_randoms_halos_expectation}) with respect to the bias function and one-point PDF\footnote{Note that if we write $B_{i}(\vs) = \int_{\DS i} d\delta_R b(\delta_R, \vs) \prob(\delta_R) / \int_{\DS i} d\delta_R \prob(\delta_R)$, then for the correlation function between density-splits $i$ and $j$ (which was also used in~\cite{paillas_cosmological_2023}, with $i=j$), we would get: $\xi_R^{i, j}(\vs) = B_i(\vs) B_j(\vs) \xi_{R}(\vs)$.}: 
\begin{equation}
    \xiRDS(\vs) = \frac{\int_{\DS} d\delta_R b(\delta_R, \vs) \prob(\delta_R)}{\int_{\DS} d\delta_R \prob(\delta_R)} \xi_R(\vs).
\label{eq:density_split_corr_bias}
\end{equation}
We note that with a linear Kaiser bias function: $b(\delta_R) = \delta_R / \sigma_R^2$, we recover equation~\eqref{eq:density_splits_gaussian_model}.

To estimate the bias function in the simulations, as the density contrast is computed on a mesh, we use the smoothed density contrast computed on all the nodes of the mesh as a sample for $\delta_R(\vr)$. To sample $\delta_R(\vr + \vs)$, we take the average of the smoothed density contrast at 3 (out of the 6, not to double count pairs) positions of the mesh separated by a distance $s$ from each node, where $s$ takes discrete values $n \times 5 \; \hMpc$, with $n$ an integer (see~\cite{codis_large-scale_2016, uhlemann_beyond_2017}). As mentioned in section~\ref{sec:cic_density_estimator}, we only work in real space here so the direction of the separation does not matter. 

Then we compute the sample average of $\delta_R(\vr + \vs)$ for each unique value of $\delta_R(\vr)$. Our bias function estimator reads, for a given separation $\vs$: 
\begin{equation}
    \widehat{b}(\delta_R, \vs) = \frac{1}{\widehat{\xi_R}(\vs)} \left[ \frac{1}{3 |\delta_R|} \sum_{\substack{i, \\ \delta_{R, i} = \delta_R}} \sum_{\substack{j=1, \\ \lVert \vx_j - \vx_i \rVert = s}}^{3} \delta_{R, j}\right]
\end{equation}
where $|\delta_R|$ denotes the number of mesh nodes where the smoothed density contrast is equal to $\delta_R$, $\delta_{R, i}$ is the value of the smoothed density contrast at 
node $i$ and $\vx_i$ denotes the positions of the mesh nodes. $\widehat{\xi_R}(s)$ is an estimator of the smoothed two-point correlation function at separation $s$ given by:
\begin{equation}
    \widehat{\xi_R}(\vs) = \langle \delta_R(\vr) \delta_R(\vr + \vs) \rangle
    = \frac{1}{3N} \sum_{i} \sum_{\substack{j=1, \\ \lVert \vx_j - \vx_i \rVert = s}}^3 \delta_{R, i} \delta_{R, j}
\label{eq:smoothed_corr_estimator}
\end{equation}
where N is the total number of mesh nodes.

\section{Model}\label{sec:model}

In this section, we derive predictions for the density-split correlation functions $\xiRDS$, based on two models for the joint PDF $\prob(\delta_{R}(\vr), \delta_{R}(\vr + \vs))$: a log-normal approximation and a model based on the LDT formalism.

\subsection{(Shifted) log-normal approximation}\label{sec:log-normal}
In this section, we assume that $(\delta_{R}(\vr), \delta_{R}(\vr + \vs))$ follows a bivariate shifted log-normal distribution~\cite{hilbert_cosmic_2011, xavier_improving_2016, friedrich_density_2018, uhlemann_it_2023}. In other words, if we make the following change of variable: 
\begin{equation}
    Y_{R} = \ln \left( 1 + \frac{\delta_{R}}{\delta_{0}} \right) + \frac{\sigmaYR^2}{2},
\end{equation}
with $\delta_{0}$ a free parameter (set by the skewness of the distribution), we assume that $Y_{R}(\vr), Y_{R}(\vr + \vs)$ follows a bivariate Gaussian distribution centered on 0, with covariance matrix:
\begin{equation}
\Sigma_{Y_R}(\vs) = 
\begin{pmatrix}
    \sigmaYR^2 & \xiYR(\vs)\\
    \xiYR(\vs) & \sigmaYR^2
\end{pmatrix}  
\end{equation}
where $\sigmaYR^2$ is the variance of $Y_{R}$, and $\xiYR(\vs) = \langle Y_{R}(\vr) Y_{R}(\vr + \vs) \rangle$. We find that $\sigmaYR^2$ and $\xiYR(\vs)$ verify~\cite{hilbert_cosmic_2011}:
\begin{equation}
    \sigmaYR^2 = \ln \left( 1 + \frac{{\sigma_R}^2}{ \delta_{0}^2} \right)
\end{equation}
and:
\begin{equation}
    \xiYR(\vs) = \ln \left( 1 + \frac{\xi_{R}(\vs)}{\delta_{0}^2} \right),
\end{equation}
where $\sigma_R$ and $\xi_{R}(\vs)$ are the variance of the smoothed density contrast and the smoothed two-point correlation function, respectively, as defined in section~\ref{sec:gaussian_density}.
$\delta_{0}$ is determined by the following relation~\cite{friedrich_density_2018}:
\begin{equation}
    \langle \delta_R^3 \rangle = \frac{3 \langle \delta_R^2 \rangle^2}{\delta_0}  + \frac{\langle \delta_R^2 \rangle^3}{\delta_0^3},
\label{eq:delta0_skewness}
\end{equation}
but in practice we fit it along with $\sigma_R^2$ to match the measured distribution.

\subsubsection{One-point PDF}
In this shifted lognormal model, the one-point PDF of $\delta_R$ then reads
\begin{equation}
    \prob(\delta_R) =
    \frac{1}{\sqrt{2 \pi} \sigmaYR (\delta_R + \delta_{0})} \exp \left[-\frac{ \left( \ln \left( 1 + \frac{\delta_R}{\delta_{0}} \right) + \frac{\sigmaYR^2}{2} \right)^2 }{2\sigmaYR^2}, \right]
\end{equation}
for $\delta_R > - \delta_{0}$ (while $\prob(\delta_R) = 0$ for $\delta_R \leq -\delta_{0}$). Additionally, we convolve this PDF with a Poisson shot noise to account for the finite number of (downsampled) particles in our simulations (see for instance \cite{repp_galaxy_2020, friedrich_density_2018}):
\begin{equation}
    \prob_{\rm SN}(N) = \int d\delta_R \prob(N|\delta_R)\prob(\delta_R)
\label{eq:one-point_pdf_shotnoise}
\end{equation}
where
\begin{equation}
    \prob(N|\delta_R) = \poisson(N, \nbar V_R (1 + \delta_R)),
\label{eq:one-point_pdf_poisson}
\end{equation}
with $N$ the number of particles within the cell, $\nbar$ the average particle number density and $V_R = 4 \pi R^3 / 3$ the volume of the spherical cell of radius $R$. Hence our final prediction for the one-point PDF of the density $\delta_R$ is
\begin{equation}
    \prob_{\rm SN}(\delta_R) = \nbar V_R \prob_{\rm SN}(N = \bar{n}V_R(1 + \delta_R)).
\label{eq:one-point_pdf_final}
\end{equation}
There are two free parameters in this prediction: $\sigmaYR$ and $\delta_0$, which are fitted from the one-point PDF by minimizing the sum of squares of the standardized residuals. The shifted log-normal (+ shot noise) prediction for the density one-point PDF is shown in figure~\ref{fig:one-point_pdf} (dotted line), along with the measurement from the 25 AbacusSummit simulations. The log-normal approximation with shot noise matches well the measurement, the error being within 1\% of the PDF in the range $\delta_R \in [-1, 3]$, although the residuals are larger than the standard deviation of the mocks. 
The effect of the convolution with Poisson shot noise is shown in figure~\ref{fig:shotnoise} (appendix~\ref{sec:shotnoise}).
To get a sense of the model precision required to scale with current galaxy surveys such as DESI, in figure~\ref{fig:one-point_pdf_lownbar} we show the comparison between the log-normal prediction and the measurement from the lower density simulations ($\nbar = 5 \times 10^{-4} \; (\Mpch)^3$), which have an effective volume comparable to that of the DESI DR1 \texttt{LRG3 + ELG1} sample, at two different smoothing radii. The agreement between the model and simulations is relatively good, mostly within one standard deviation of the mocks for both smoothing radii.

An important remark is that we fitted both $\sigma_R$ and $\delta_0$ (by minimizing the sum of the squares of the standardized residuals between the log-normal model and the simulation measurements). We could also have used the prediction for the skewness $\langle \delta_R^3 \rangle$ from perturbation theory to fix $\delta_0$ through equation~\eqref{eq:delta0_skewness}. This would be a fairer comparison to the LDT model presented below (see section~\ref{sec:ldt}), which predicts the skewness (with the same prediction as tree-order perturbation theory since it is based on spherical collapse).

\subsubsection{Two-point PDF}

The two-point PDF of the bivariate log-normal distribution reads:
\begin{equation}
    \prob(\delta_R, \delta_R^{\prime}) =
    \frac{1}{2 \pi \sigmaYR^2 \sqrt{1 - \rhoYR^2} (\delta_R + \delta_{0}) (\delta_R^{\prime} + \delta_{0})} \exp \left[
    -\frac{{Y_R}^2 + {Y_R^{\prime}}^2 - 2 \rho_{Y, R} Y_R Y_R^{\prime}}
    {2 \sigmaYR^2 (1 - \rhoYR^2)} 
    \right]
\label{eq:log-normal_two_point_pdf}
\end{equation}
for $\delta_R, \delta_R^{\prime} > - \delta_{0}$ (and $\prob(\delta_R, \delta_R^{\prime}) = 0$ otherwise), with $\rhoYR = \xiYR(\vs)/\sigmaYR^2$, and where for simplification we noted $\delta_R = \delta_R(\vr)$ and $\delta_R^{\prime} = \delta_R(\vr + \vs)$. 

To account for shot noise, similarly to the 1D case, we convolve this PDF with a Poisson distribution:
\begin{equation}
    \prob_{\rm SN}(N, N^{\prime}) 
    = \int d\delta_R \int d\delta_R^{\prime} \prob(N|\delta_R) \prob(N^{\prime}|\delta_R^{\prime}) \prob(\delta_R, \delta_R^{\prime}),
\end{equation}
and then:
\begin{equation}
    \prob_{\rm SN}(\delta_R, \delta_R^{\prime}) 
    = (\nbar V_R)^2 \prob_{\rm SN}(N = \bar{n}V_R(1 + \delta_R), N^{\prime} = \bar{n}V_R(1 + \delta_R^{\prime})).
\label{eq:two_point_pdf_shotnoise}    
\end{equation}

\subsubsection{Bias function}
Starting from the definition in equation~\eqref{eq:bias_function}, and injecting in equation~\eqref{eq:log-normal_two_point_pdf}, we obtain the following expression for the bias function in the log-normal approximation:
\begin{equation}
    b(\delta_R, \vs) = \frac{\delta_0}{\xi_R(\vs)}  
    \left[ \exp \left( \frac{\xiYR(\vs)}{\sigmaYR^2} Y_R - \frac{\xiYR^2(\vs)}{2 \sigmaYR^2} \right) - 1 \right].
\label{eq:bias_function_log-normal}
\end{equation}
In the large separation limit, i.e. when $\xiYR(\vs) \ll \sigmaYR^2 $ and $\xi_{R}(\vs) \ll \delta_0^2 $, we find the result from~\cite{uhlemann_it_2023} (equation (30)):
\begin{equation}
    b(\delta_R, \vs) \sim \frac{Y_R}{\delta_0 \sigmaYR^2}.
\end{equation}
To compute prediction of the bias function from equation~\eqref{eq:bias_function_log-normal} numerically, we estimate $\xi_R(\vs)$ directly from the mocks, using the estimator from equation~\eqref{eq:smoothed_corr_estimator}, while $\sigmaYR$ and $\delta_0$ are fixed to the best-fit values from the one-point PDF.

To get the bias function prediction with shot noise, we combine equations~\eqref{eq:bias_peak_background} and  \eqref{eq:two_point_pdf_shotnoise} and obtain:
\begin{equation}
    b_{\rm SN}(\xdeltaR, \vs) = \nbar V_R \frac{\int d\delta_R \prob(N=\nbar V_R (1 + \xdeltaR)|\delta_R)\prob(\delta_R)b(\delta_R, \vs)}{\prob_{\rm SN}(\xdeltaR)}.
\label{eq:bias_function_shotnoise}
\end{equation}
The effect of this convolution with a Poisson distribution is shown in the right panel of figure~\ref{fig:shotnoise} (appendix~\ref{sec:shotnoise}). Figure~\ref{fig:bias_function} presents the log-normal prediction of the bias function, including shot noise (dotted line). The log-normal approximation tends to under-predict the bias for extreme values of $\delta_R$, and over-predict it for $\delta_R$ around 0. Figure~\ref{fig:bias_function_lownbar} shows the bias function prediction for the lower density simulations, for two different smoothing radii, at separation $s = 40 \; \hMpc$. The agreement is a bit better in the lower density case with $R = 10 \; \hMpc$, (left panel of figure~\ref{fig:bias_function_lownbar}) compared to the higher density at the same smoothing radius (right panel of figure~\ref{fig:bias_function}), because the former is dominated by shot noise.

\subsubsection{Density-split correlation function}
Assuming $\delta_R$ follows a shifted log-normal distribution, from the definition of $|\DS|$ in equation~\eqref{eq:DS_def}, we have
\begin{equation}
    |\DS| = \frac{1}{2} 
    \left[ \mathrm{erf} \left(\frac{Y}{\sqrt{2}\sigmaYR}\right) 
    \right]_{\ln\left(1 + \frac{\delta_{1}}{\delta_{0}}\right) + \sigmaYR^2 / 2 }^{\ln\left(1 + \frac{\delta_{2}}{\delta_{0}}\right) + \sigmaYR^2 / 2}.
    \label{eq:n_DS_avg_smooth_randoms_shifted_log-normal}
\end{equation}
Injecting the bivariate log-normal distribution from equation~\eqref{eq:log-normal_two_point_pdf} in the definition of the density-split correlation functions in equation~\eqref{eq:cross_corr_randoms_halos}, we get
\begin{equation}
\begin{split}
    |\DS| \left( 1 + \xiRDS(\vs) \right)
    & = \frac{1 - \delta_{0}}{2} \left[ \mathrm{erf} \left(\frac{Y}{\sqrt{2}\sigmaYR}\right) 
    \right]_{\ln\left(1 + \frac{\delta_{1}}{\delta_{0}}\right) + \sigmaYR^2 / 2}
    ^{\ln\left(1 + \frac{\delta_{2}}{\delta_{0}}\right) + \sigmaYR^2 / 2} \\
    & + \frac{\delta_{0}}{2} 
    \left[ \mathrm{erf} \left(\frac{Y}{\sqrt{2}\sigmaYR}\right) 
    \right]_{\ln\left(1 + \frac{\delta_{1}}{\delta_{0, R}}\right) + \sigmaYR^2 / 2 - \xiYR(\vs)}
    ^{\ln\left(1 + \frac{\delta_{2}}{\delta_{0, R}}\right) + \sigmaYR^2 / 2 - \xiYR(\vs)}
    \label{eq:n_DS_crosscorr_smooth_randoms_shifted_log-normal_1}
\end{split}
\end{equation}
such that, with $\delta_1$, $\delta_2$ the DS boundaries, the log-normal prediction for the density-split correlation function, neglecting shot noise, reads
\begin{equation}
    \xiRDS(\vs) = \delta_{0} \left(
    \frac{\left[ \mathrm{erf} \left(\frac{Y}{\sqrt{2}\sigmaYR}\right) 
    \right]_{\ln\left(1 + \frac{\delta_{1}}{\delta_{0}}\right) + \sigmaYR^2 / 2 - \xiYR(\vs)}
    ^{\ln\left(1 + \frac{\delta_{2}}{\delta_{0}}\right) + \sigmaYR^2 / 2 - \xiYR(\vs)}}
    {\left[ \mathrm{erf} \left(\frac{Y}{\sqrt{2}\sigmaYR}\right) 
    \right]_{\ln\left(1 + \frac{\delta_{1}}{\delta_{0}}\right) + \sigmaYR^2 / 2 }^{\ln\left(1 + \frac{\delta_{2}}{\delta_{0}}\right) + \sigmaYR^2 / 2}}
    -1 \right).
    \label{eq:xi_R_DS_randoms_shifted_log-normal}
\end{equation}

To add shot noise to the prediction, we perform a numerical integration of equation~\eqref{eq:density_split_corr_bias}, where we use the model with shot noise for $b(\delta_R, \vs)$ and $\prob(\delta_R)$. Here again, we use the values of $\sigmaYR$ and $\delta_0$ fitted to the one-point PDF, while for $\xiYR$ we measure the average smoothed two-point correlation function $\widehat{\xi_R}(\vs)$ from the 25 simulations, using Landy-Szalay estimator, and take $\xiYR(\vs) = \ln(1 + \widehat{\xi_R}(\vs)/\delta_0^2) $. Figure~\ref{fig:density_splits_comparison} shows the log-normal prediction for the three density-split correlation functions $\xiRDS$ along with the average measurement from simulations. The density-splits' edges in the model ($\delta_1$, $\delta_2$ in equation~\eqref{eq:xi_R_DS_randoms_shifted_log-normal}) are the same as those used in the measurements from simulations. The bottom panel of figure~\ref{fig:density_splits_comparison} shows the difference between the model and the simulations, divided by the standard deviation of the simulations. The pink area displays the ratio of the standard deviation of the lower density simulations, which are representative of DESI DR1 \texttt{LRG3 + ELG1} sample, to the standard deviation of the higher density simulations. The log-normal model is in agreement with the simulations, but it is not at the level of the standard deviation of the 25 simulations for the lower density quantiles ($\DS$ 1 and 2). The same conclusion applies with five density-splits (figure~\ref{fig:five_density_splits_comparison}), where the agreement is good relative to the correlation function values, but not sufficient when comparing with the standard deviation of the mocks. The agreement is better for the higher density quantile, which is explained by the bias function being better modeled for high densities, with respect to the standard deviation of the measurements (see figure~\ref{fig:bias_function}).

\subsection{Large Deviation Theory}\label{sec:ldt}

In this section, we recall the LDT framework and its predictions for the count-in-cell PDF and the bias function in the large separation limit. The density PDF is said to satisfy a \textit{large deviation principle} (LDP) if the following limit, called \textit{rate function}, exists~\cite{bernardeau_large_2016, uhlemann_back_2016}:
\begin{equation}
    \psi(\delta_R) = - \lim_{\sigma_R^2 \rightarrow{} 0} \sigma_R^2 \log \prob(\delta_R).
\label{eq:ldp}
\end{equation}
The rate function $\psi(\rho_R)$ governs the exponential decay of the density PDF as the variance of the smoothed density contrast $\sigma_R^2$ -- the \textit{driving parameter} -- goes to zero:
\begin{equation}
    \prob(\delta_R) \underset{\sigma_R^2 \rightarrow{} 0}{\propto} \exp{ \left[ - \frac{\psi(\delta_R)}{\sigma_R^2} \right]}.
\label{eq:ldp_pdf}
\end{equation}
In particular, in the case of a Gaussian density PDF, for example if we assume a Gaussian initial density contrast $\delta_{L, \; R}$, the rate function is simply $\psi(\delta_{L, \; R}) = \delta_{L, \; R}^2 / 2$ where the subscript $L$ stands for "linear". We also define the decay-rate function, which is the ratio of the rate function and the variance: $\Psi(\delta_{L, \; R}) = \psi(\delta_{L, \; R})/\sigma_{L, \; R}^2$. 

A consequence of the LDP, called \textit{contraction principle}, is that if $\tau$ is a random variable satisfying the LDP, then for any continuous mapping $\mathcal{F}$ relating $\tau$ to $\delta_R$, $\delta_R$ also satisfies the LDP, and we have~\cite{bernardeau_large_2016}:
\begin{equation}
    \psi(\delta_R) = \inf_{\tau | \delta_R = \mathcal{F}(\tau)} \psi_{\tau}(\tau).
\label{eq:contraction_principle}
\end{equation}
In particular, for spherically symmetric configurations, it is natural to assume that the most likely dynamics relating the initial field and the late-time density (and therefore realising the infimum in equation~\eqref{eq:contraction_principle}) should also respect this symmetry hence be the spherical collapse dynamics. In this context, one can
relate the evolved, non-linear density $\rho_R = 1 + \delta_R$ to the initial density $\tau$ in a cell at the same location that contains the same mass, i.e. with radius $r = R\rho_R^{1/3}$. Although solving the spherical collapse dynamics often requires numerical computations, an accurate approximation can be simply written as:
\begin{equation}
    \rho_R \simeq \frac{1}{(1 - \tau/\nu)^\nu}
\label{eq:spherical_collapse}
\end{equation}
with $\nu = \frac{21}{13}$~\cite{bernardeau_large_2016}. Then, assuming initial Gaussian fluctuations, the late-time rate function is simply given by
\begin{equation}
    \psi(\delta_R) = \frac{\sigma_{L, \; R}^2}{2 \sigma_{L, \; r}^2} \tau(\rho_R)^2.
\label{eq:rate_function}
\end{equation}
To compute the PDF of the late-time, non-linear density contrast $\delta_R$, we first compute its cumulant generating funtion. The scaled cumulant generating function of $\delta_R$, $\varphi(\lambda)$, where $\lambda$ is a dimensionless parameter, is related to the cumulant generating function $\phi$ through
\begin{equation}
    \varphi(\lambda) = \lim_{\sigma_R^2 \rightarrow{} 0} \sigma_{R}^2\, \phi \left( \frac{\lambda}{\sigma_R^2} \right)
\end{equation}
and can be obtained as the Legendre-Fenchel transform of the rate function~\cite{bernardeau_large_2016}
\begin{equation}
    \varphi(\lambda) = \sup_{\delta_R} \left[ \lambda \delta_R - \psi(\delta_R) \right].
\end{equation}
Now, to get the cumulant generating function for any variance $\sigma_R^2$, we use the following prescription which has been shown to give accurate results in the mildly non-linear regime \cite{bernardeau_statistics_2014}:
\begin{equation}
    \phi(\lambda) \simeq \frac{1}{\sigma_R^2} \varphi(\lambda \sigma_R^2).
\end{equation}
The one-point PDF of the density is then given by the inverse Laplace transform of the moment generating function:
\begin{equation}
    \prob(\rho_R) = \int_{-i\infty}^{+i\infty} \frac{d\lambda}{2\pi i} \exp \left[ -\lambda \rho_R + \phi(\lambda) \right].
\label{eq:laplace}
\end{equation}

\subsubsection{One-point PDF}

The Laplace transform in equation~\eqref{eq:laplace} can be approximated using the saddle-point approximation, assuming that $\psi^{\prime \prime} (\rho_R) > 0$. To avoid a critical point where $\psi^{\prime \prime} (\rho_{R, c}) = 0$, \cite{uhlemann_back_2016} proposed to perform a change of variable $\mu = \log \rho_R$ to compute the density PDF. This approach gives the following LDT prediction for the one-point PDF of the count-in-cell density in spheres of radius $R$:
\begin{equation}
    \prob(\rho_R) = \sqrt{\frac{\psi^{\prime \prime} (\rho_R) + \psi^{\prime}(\rho_R)/\rho_R}{2 \pi \sigma_{R, \; \eff}^2}} \exp \left( - \frac{\psi (\rho_R)}{\sigma_{R, \; \eff}^2} \right),
\label{eq:one-point_pdf}
\end{equation}
where primed quantities denote the derivative with respect to $\rho_R$, and $\sigma_{R, \; \eff}$ is defined below (equation~\eqref{eq:effective_variance}). As equation~\eqref{eq:one-point_pdf} is an approximation to the exact PDF of $\rho_R$, we need to ensure that it is properly normalized and that $\langle \rho_R \rangle = 1$ \footnote{To be more precise, the constraint on the mean of the resulting density field has to be added since applying LDT to the log-density predicts all $(n\leq3)$-order cumulants but not the mean. To get the right value of the mean of the non-linearly evolved log-density field, one hence has to solve for the condition that the mean density contrast is zero.}. Hence, we use the following expression for the one-point count-in-cell PDF~\cite{uhlemann_back_2016, uhlemann_beyond_2017}:
\begin{equation}
    \widehat{\prob}(\rho_R) = \prob \left( \rho_R \times \frac{\langle \rho_R \rangle}{\langle 1 \rangle} \right) \times \frac{\langle \rho_R \rangle}{\langle 1 \rangle^2}
\label{eq:one-point_pdf_normalized}
\end{equation}
where $\langle 1 \rangle = \int d\rho_R \prob(\rho_R)$ and $\langle \rho_R \rangle = \int d\rho_R \rho_R \prob(\rho_R)$. Therefore, to ensure the correct variance for the PDF prediction, $\sigma_{R, \; \eff}$ is defined through the following equation:
\begin{equation}
    1 + \sigma_R^2 = \frac{\langle 1 \rangle}{\langle \rho_R \rangle^2} \langle \rho_R^2 \rangle
\label{eq:effective_variance}
\end{equation}
where $\sigma_R$ is fitted to the measured PDF\footnote{The variance $\sigma_R^2$, which is not predicted by LDT, could be either fitted or predicted by any emulator of the non-linear matter power spectrum.} and $\langle \rho_R^2 \rangle = \int d\rho_R \rho_R^2 \prob(\rho_R)$.
Similarly to the log-normal model, we convolve this PDF with a Poissonian shot noise to account for the finite number of particles in the simulations (\cref{eq:one-point_pdf_shotnoise,eq:one-point_pdf_poisson,eq:one-point_pdf_final}). The left panel of figure~\ref{fig:shotnoise} (appendix~\ref{sec:shotnoise}) shows the effect of the convolution with Poisson shot noise on the LDT-only distribution.

\begin{figure}
\centering 
\includegraphics[scale=0.8]{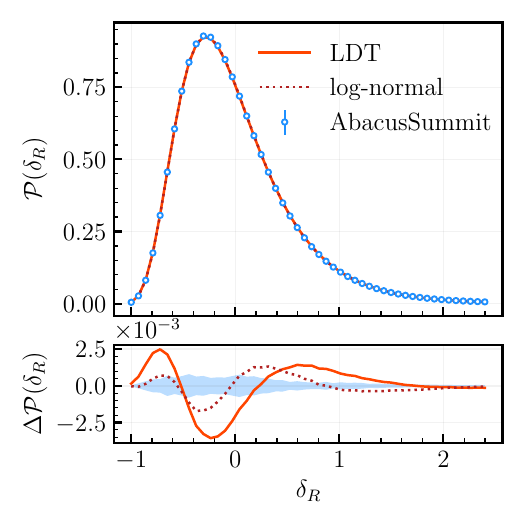}
\caption{\textit{Top:} average density PDF in 25 AbacusSummit dark matter particles simulations with $\nbar = 0.0034 \; (\Mpch)^3$ and $R = 10 \; \hMpc$, compared with log-normal and LDT models. Poisson shot noise is included in both models. \textit{Bottom:} difference between each model and the simulations. The blue area represents the standard deviation of the 25 mocks.}
\label{fig:one-point_pdf}
\end{figure}

Our LDT model is implemented in Python, adapted from the Mathematica package \texttt{LSSFast}~\cite{codis_encircling_2016}. We compare this prediction with the measurement from the 25 AbacusSummit simulations with $\nbar = 0.0034 \; (\Mpch)^3$ in figure~\ref{fig:one-point_pdf}.  LDT prediction is slightly less accurate than the log-normal approximation (although below the percent precision on the PDF), but with only one free parameter ($\sigma_R$) -- compared to two free parameters in the log-normal distribution.  The skewness $\langle \delta_R^3 \rangle$ of the LDT (+ shot noise) model is 0.234, which is below the one we measure from the simulations: $0.241 \pm 0.001$. This not surprising as the LDT prediction for the skewness is that of perturbation theory at tree order, while non-linearities tend to generate a larger skewness~\cite{gaztanaga_hierarchical_1995}.

\begin{figure}
\begin{subfigure}{0.5\textwidth}
\includegraphics[scale=0.8]{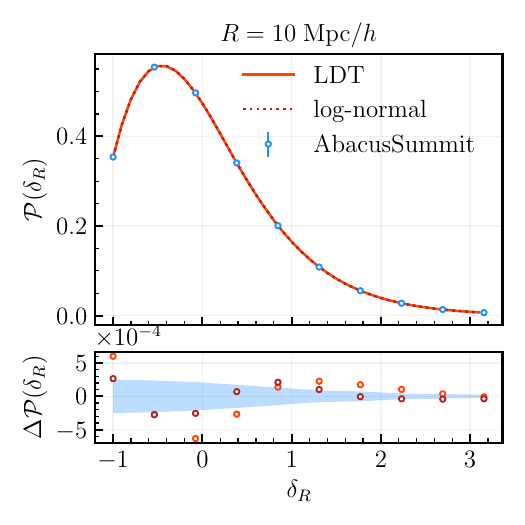}
\end{subfigure}
\begin{subfigure}{0.5\textwidth}
\includegraphics[scale=0.8]{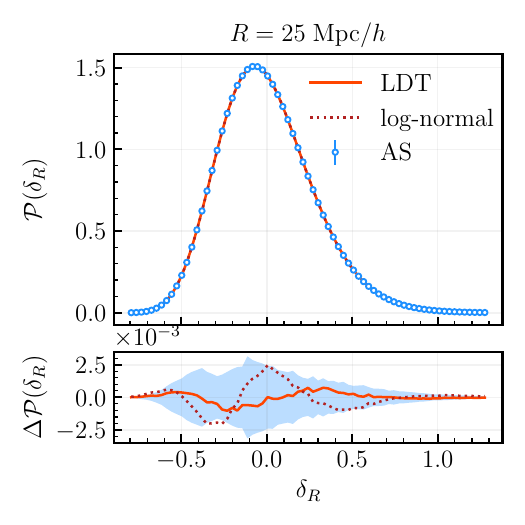}
\end{subfigure}
\caption{Average density PDF in 25 AbacusSummit (AS on the right panel) dark matter particles simulations with low density ($\nbar = 5 \times 10^{-4} \; (\Mpch)^3$), compared with log-normal and LDT models. Poisson shot noise is included in both models. Note that because of shot noise, the models give predictions for discrete values of $\delta_R$ (the same as the data points). In the left panel ($R = 10 \; \hMpc$), since the points are sparse, we display a cubic spline interpolation of the models to improve the figure's readability. Bottom panel shows the difference between each model and the simulations. Blue area shows the standard deviation of the 25 mocks. In the bottom left panel, red circles represent the LDT residuals, while dark brown circles represent the log-normal residuals.} \textit{Left}: $R = 10 \; \hMpc$.  \textit{Right}: $R = 25 \; \hMpc$.
\label{fig:one-point_pdf_lownbar}
\end{figure}

In figure~\ref{fig:one-point_pdf_lownbar} we show the comparison between the LDT prediction and the measurement from the lower density simulations ($\nbar = 5 \times 10^{-4} \; (\Mpch)^3$), which have an effective volume comparable to that of the DESI DR1 \texttt{LRG3 + ELG1} sample. For a smoothing radius $R = 10 \; \hMpc$, the residuals between the model and the simulations are still larger than the standard deviation among the mocks, although contained within $3 \sigma$. The agreement is well below the standard deviation of the mocks if we increase the smoothing radius to $R = 25 \; \hMpc$. Although the log-normal model is in better agreement with the mocks than the LDT model at $R = 10 \; \hMpc$ case, we see that the LDT model performs much better for a larger smoothing radius.

\subsubsection{Bias function}
For low densities and in the large separation limit $s \gg R$, the bias function defined in equation~\eqref{eq:bias_function} can be predicted with~\cite{codis_encircling_2016, codis_large-scale_2016}:
\begin{equation}
    b(\rho_R) = \frac{\tau(\rho_R)}{\sigma_{r, L}^2}
\end{equation}
which does not depend on $\vs$. To ensure the normalization of the density two-point PDF in equation~\eqref{eq:bias_peak_background} and the definition of $\xi_R$, the bias function must verify the following two relations~\cite{uhlemann_beyond_2017}:
\begin{equation}
    \langle b(\rho_R) \rangle = \int d\rho_R b(\rho_R) \widehat{\prob}(\rho_R) = 0,
\end{equation}
\begin{equation}
    \langle \rho_R b(\rho_R) \rangle = \int d\rho_R \rho_R b(\rho_R) \widehat{\prob}(\rho_R) = 1.
\label{eq:bias_normalization}
\end{equation}
Hence the bias function prediction without shot noise is:
\begin{equation}
    \widehat{b}(\rho_R) = \frac{b(\rho_R) - \langle b(\rho_R) \rangle}{\langle \rho_R b(\rho_R) \rangle - \langle b(\rho_R) \rangle}.
\end{equation}
As for the log-normal model, we convolve this bias model with a Poisson shot noise (\cref{eq:bias_function_shotnoise}). The effect of shot noise is shown in appendix~\ref{sec:shotnoise}.

\begin{figure}
\begin{subfigure}{0.5\textwidth}
\includegraphics[scale=0.8]{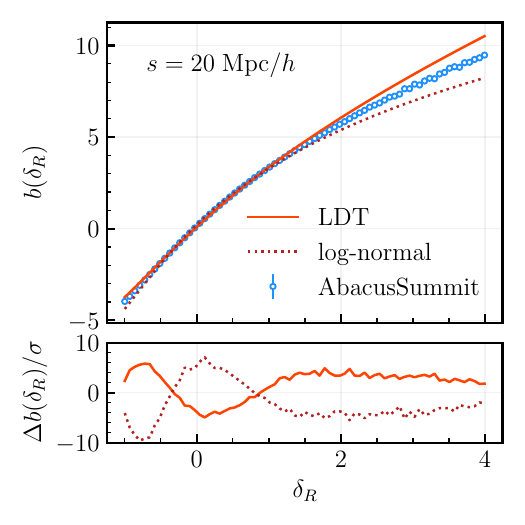}
\end{subfigure}
\begin{subfigure}{0.5\textwidth}
\includegraphics[scale=0.8]{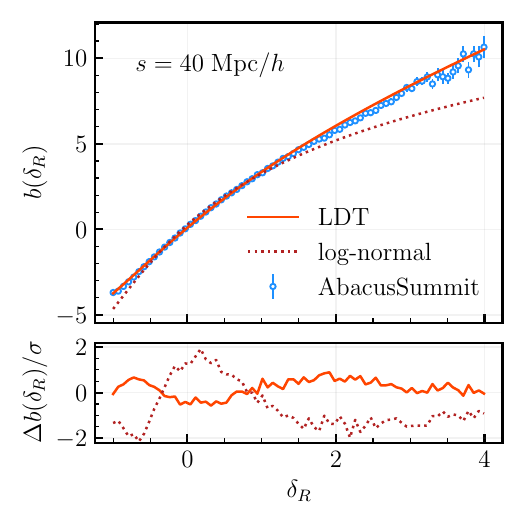}
\end{subfigure}
\caption{\textit{Top:} Average bias function in 25 AbacusSummit dark matter particles simulations with $R = 10 \; \hMpc$, at separations $s = 20 \; \hMpc$ (left) and $s = 40 \; \hMpc$ (right), with $\nbar = 0.0034 \; (\Mpch)^3$. Error bars show the standard deviation over the 25 mocks divided by 5. Solid (dotted) line shows the LDT (log-normal) model prediction, including Poisson shot noise. The LDT model is the large-separation approximation, so it is the same in both panels. \textit{Bottom:} Difference between model and mocks divided by the standard deviation of the 25 mocks.}
\label{fig:bias_function}
\end{figure}

\begin{figure}
\begin{subfigure}{0.5\textwidth}
\includegraphics[scale=0.8]{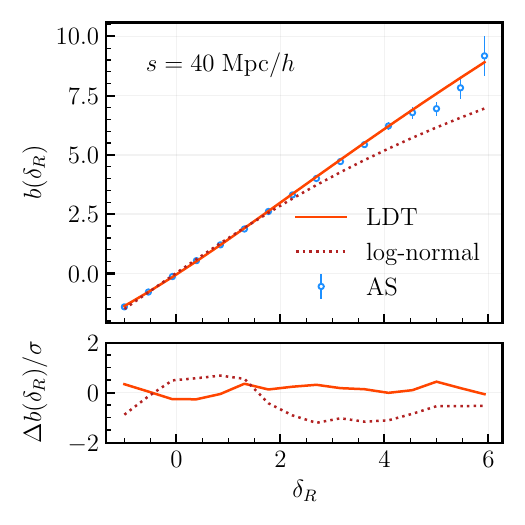}
\end{subfigure}
\begin{subfigure}{0.5\textwidth}
\includegraphics[scale=0.8]{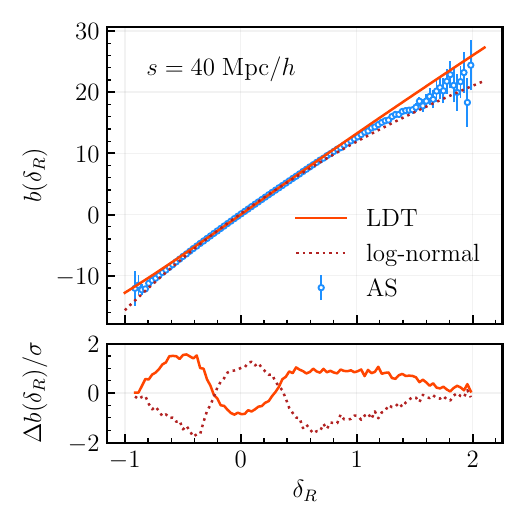}
\end{subfigure}
\caption{\textit{Top:} Average bias function in 25 AbacusSummit (AS) low density ($\nbar = 5 \times 10^{-4} \; (\Mpch)^3$) simulations with $R = 10 \; \hMpc$ (left) and $R = 25 \; \hMpc$ (right). Error bars show the standard deviation over the 25 mocks divided by 5. Solid (dotted) line shows the LDT (log-normal) model prediction, including Poisson shot noise. \textit{Bottom:} Difference between model and mocks divided by the standard deviation of the 25 mocks.
}
\label{fig:bias_function_lownbar}
\end{figure}

The only free parameter here is $\sigma_R$, which we fix to the best-fit value from the one-point PDF. Figure~\ref{fig:bias_function} presents this bias prediction along with the measured bias function in the simulations at separations $s = 20 \; \hMpc$ and $s = 40 \; \hMpc$. The LDT model over-predicts the bias function at small separation for large absolute values of $\delta_R$, and under-predicts it for $\delta_R$ close to 0. The LDT model performs better than the log-normal approximation at large separations with respect to the smoothing radius $R = 10 \hMpc$. The agreement is excellent at $s = 40 \; \hMpc$, where the residuals between the model and the simulations fall within one standard deviation of the mocks. This behaviour of the LDT prediction is expected since the large separation expansion can only hold when the two separated spheres at least do not overlap, i.e. for $s>2R$. Figure~\ref{fig:bias_function_lownbar} displays the LDT predictions with measurements from the low density simulations, for $R = 10 \hMpc$ and $R = 25 \hMpc$. For both smoothing radii, the agreement between the LDT model and the simulations is very good, even though $s$ is not large with respect to $R = 25 \hMpc$.

\subsubsection{Density-split correlation function}

The density-split correlation functions are written in terms of the bias function in equation~\eqref{eq:density_split_corr_bias}. Including shot noise and renormalization, this yields:
\begin{equation}
    \widehat{\xiRDS}(\vs) = \frac{\int_{\DS} d\delta_R \widehat{b}_{\rm SN}(\delta_R) \widehat{\prob}_{\rm SN}(\delta_R)}{\int_{\DS} d\delta_R \widehat{\prob}_{\rm SN}(\delta_R)} \widehat{\xi_R}(\vs).
\end{equation}
Thus, in the large separation limit, the density-split correlation function is linearly related to the smoothed two-point correlation function, similarly to the Gaussian case (\cref{eq:density_splits_gaussian_model}). The difference with respect to the Gaussian case comes from the bias function, which is not linear anymore. However, the bias factor is still independent of the separation, for sufficiently large separations with respect to $R$.

\begin{figure}
\begin{subfigure}{0.5\textwidth}
\includegraphics[scale=0.8]{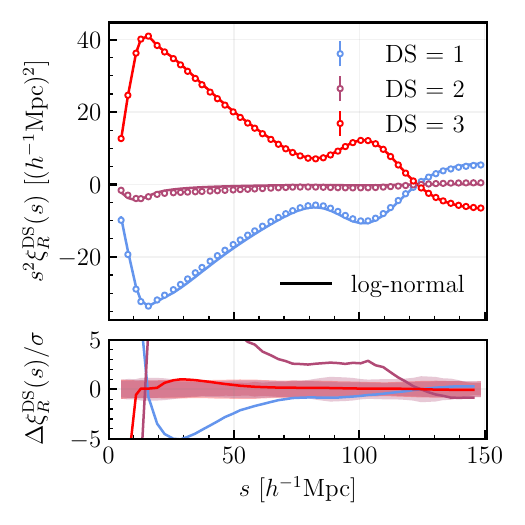}
\end{subfigure}
\begin{subfigure}{0.5\textwidth}
\includegraphics[scale=0.8]{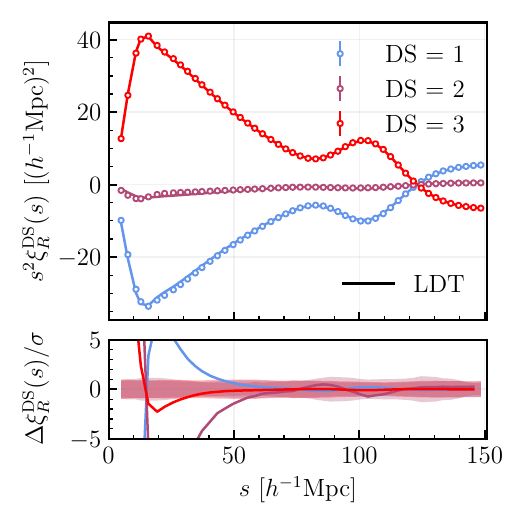}
\end{subfigure}
\caption{Average density-split correlation functions of the 25 simulations with $\nbar = 0.0034 \; (\Mpch)^3$ and $R = 10 \; \hMpc$ (circles). The left panel shows the log-normal prediction, while the right panel shows the LDT prediction. Shot noise is included in both models. Bottom panels show the difference between model and simulations, divided by the standard deviation over the 25 mocks. The shaded areas represent the standard deviation from the low density simulations ($\nbar = 5 \times 10^{-4} \; (\Mpch)^3$) divided by the standard deviation from the high density simulations ($\nbar = 0.0034 \; (\Mpch)^3$), for the three density-splits.}
\label{fig:density_splits_comparison}
\end{figure}

\begin{figure}
\begin{subfigure}{0.5\textwidth}
\includegraphics[scale=0.8]{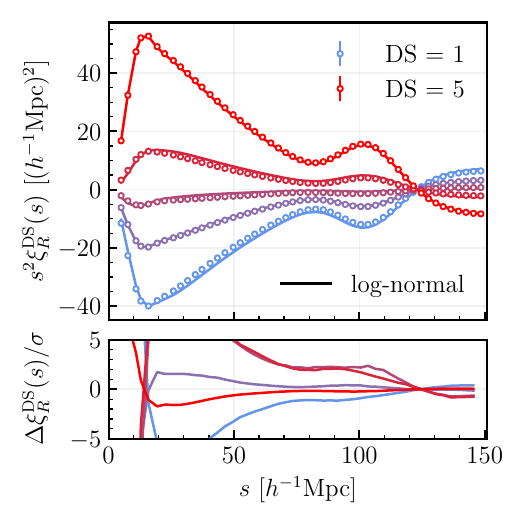}
\end{subfigure}
\begin{subfigure}{0.5\textwidth}
\includegraphics[scale=0.8]{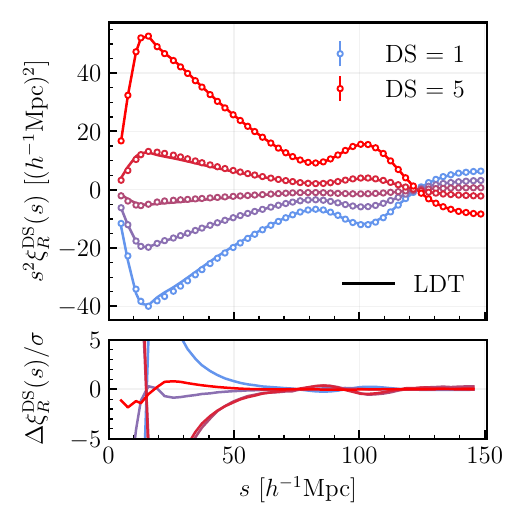}
\end{subfigure}
\caption{Same as figure~\ref{fig:density_splits_comparison} but with 5 density-splits.}
\label{fig:five_density_splits_comparison}
\end{figure}

Again, we fix the value of $\sigma_R$ to the best-fit value from the one-point PDF, and take the average smoothed two-point correlation function $\widehat{\xi_R}(\vs)$ of the mocks, computed with \texttt{pycorr} and the Landy-Szalay estimator. The right panel of figure~\ref{fig:density_splits_comparison} presents the LDT density-split model compared to the simulations. $\widehat{\xi_R}$ in the model is taken as the average smoothed correlation function of the mocks, computed with \texttt{pycorr}. The bottom panel shows the difference between the model and mocks divided by the standard deviation of the mocks. The shaded area shows the standard deviation from the lower density mocks ($\nbar = 5 \times 10^{-4} \; (\Mpch)^3$) divided by the standard deviation from the higher density simulations ($\nbar = 0.0034 \; (\Mpch)^3$) for comparison. The model agrees very well with the simulations, down to $s = 50 \; \hMpc$. The mismatch at lower scales is expected, since the LDT model for the two-point PDF is valid only in the large separation regime, for $s \gg R$. To assess the robustness of the model, we show in figure~\ref{fig:five_density_splits_comparison} the log-normal and LDT predictions with five density-splits. We see that the residuals are similar to those of the three density-splits for both models. We also note that in general, the higher density quantile tends to be better modeled than the other ones, which is because the bias function is better captured by both models at high densities.

\begin{figure}
\begin{subfigure}{0.5\textwidth}
\includegraphics[scale=0.8]{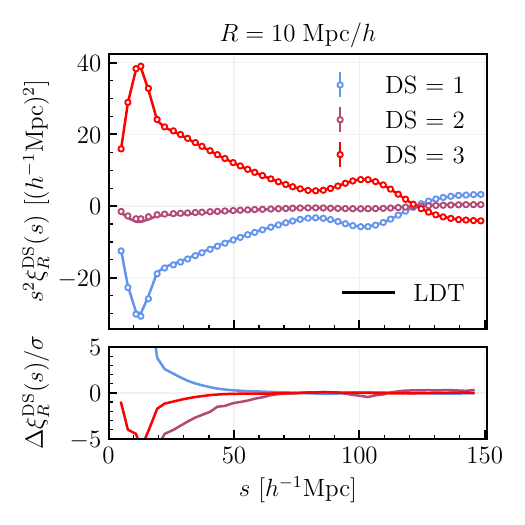}
\end{subfigure}
\begin{subfigure}{0.5\textwidth}
\includegraphics[scale=0.8]{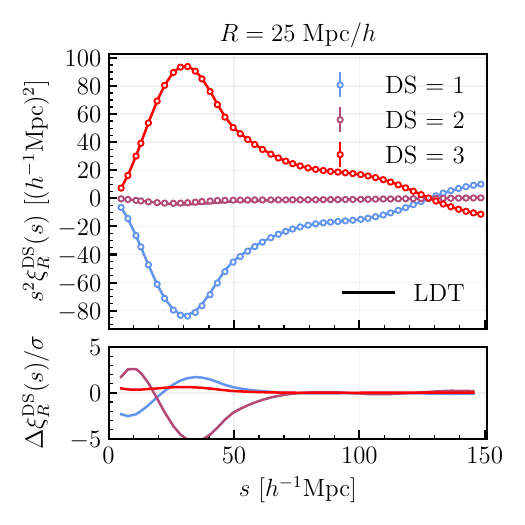}
\end{subfigure}
\caption{Average density-split correlation functions of the 25 low density ($\nbar = 5 \times 10^{-4} \; (\Mpch)^3$) simulations (circles), with $R = 10 \; \hMpc$ (left) and $R = 25 \; \hMpc$ (right). Bottom panels show the difference between the LDT model (with shot noise included) and simulations, divided by the standard deviation over the 25 mocks.}
\label{fig:density_splits_comparison_lownbar}
\end{figure}

Figure~\ref{fig:density_splits_comparison_lownbar} shows the comparison between the LDT model and the simulations with low average density with three density-splits. We see that the model performs well for separations $s \gtrsim 50 \; \hMpc$ for both smoothing radii.

We note that for those large separations, the LDT model outperforms the log-normal model, despite having fewer degrees of freedom (one with respect to two). Additionally, this one degree of freedom can be reduced to zero if we use a power spectrum emulator to predict the non-linear variance $\sigma_R^2$. Thus this framework allows for an accurate modeling of the density-splits on large scale. It provides theoretical insights into the information content of density-split clustering, complementary to commonly used simulation-based models. Moreover, it can be used as an initial guess to enhance the accuracy of such emulator. 

However, to be able to apply our approach with real survey data, further developments are required. First, galaxy surveys observe galaxies, which are biased tracers of matter. In section~\ref{sec:biased_tracers}, we explore avenues to extend this model to biased tracers by incorporating a quadratic Eulerian bias or a Gaussian Lagrangian bias, and super-Poisson shot noise, as investigated in~\cite{friedrich_pdf_2022, stucker_gaussian_2024, gould_cosmology_2024}. These advances demonstrate encouraging progress in adapting the model to galaxies. Second, we have worked in real space only. In practice, spectroscopic galaxy surveys measure the galaxy radial positions by translating their redshift (impacted by their peculiar velocities) into a distance, which leads to so-called redshift-space distortions. In particular, \cite{paillas_constraining_2023} showed that identifying density-split membership in redshift space rather that real space led to very pronounced distortions in the density-split correlation functions, especially in the quadrupole. One should account for these redshift-space distortions in both the one-point PDF and the bias function in order to be able to model density-splits in redshift space, but this extension is beyond the scope of this paper.

\section{Biased tracers}\label{sec:biased_tracers}

In this section, we build upon previous work on the modeling of halo and galaxy bias \cite{friedrich_pdf_2022, stucker_gaussian_2024, gould_cosmology_2024} to extend our density-split model to biased tracers of matter. To this end, we use the 25 AbacusSummit dark matter simulations at redshift $z = 0.8$ with volume $(2 \; \hGpc)^3$, and populate them with emission line galaxies (ELGs) following the best-fit halo occupation distribution (HOD) model from~\cite{rocher_desi_2023}, with a target density of $\nbar_g = 0.002 \; (\Mpch)^3$. We generate these ELG simulations using the python package \texttt{HODDIES}\footnote{\url{https://github.com/antoine-rocher/HODDIES}}. In this section, the smoothing radius is $R = 10 \; \hMpc$ in all figures.

To model the galaxy one-point PDF, we follow~\cite{friedrich_pdf_2022, gould_cosmology_2024} and write:
\begin{equation}
    \prob_{g}(N_g) = \nbar_g V_R \int d\delta_{R, m} \prob(N_g | \delta_{R, m}) \prob_m(\delta_{R, m})
\end{equation}
where
\begin{align}
\mathcal{P}\left(N_{g} \mid \delta_{R, m}\right) 
&=\frac{1}{\alpha\left(\delta_{R, m}\right)} \exp \left(-\frac{\langle N_{g}|\delta_{R, m}\rangle}{\alpha\left(\delta_{R, m}\right)}\right) \\
& \times\left[\Gamma\left(\frac{N_{g}}{\alpha\left(\delta_{R, m}\right)}+1\right)\right]^{-1}\left(\frac{\langle N_{g}|\delta_{R, m}\rangle}{\alpha\left(\delta_{R, m}\right)}\right)^{\frac{N_{g}}{\alpha\left(\delta_{R, m}\right)}}
\end{align}
with $N_g = \nbar_g V_R (1 + \delta_{R, g})$ the number of galaxies in spheres of radius $R$. $\langle N_{g}|\delta_{R, m}\rangle$ is the conditional mean of $N_{g}$ given $\delta_{R, m}$, and $\alpha(\delta_{R, m})$ is the ratio of the conditional variance of $N_g$ with respect to its conditional mean:
\begin{equation}
    \alpha(\delta_{R, m}) = \frac{\langle N_{g}^2 | \delta_{R, m} \rangle_c}{\langle N_g | \delta_{R, m} \rangle} = N_g \frac{\langle \delta_{R,g}^2 | \delta_{R, m} \rangle}{(1 + \langle \delta_{R,g} | \delta_{R, m} \rangle)}. 
\end{equation}
$\alpha(\delta_{R, m})$ characterizes the scatter of the galaxy counts distribution around its expectation value, namely shot noise. Note that when $\alpha(\delta_{R, m}) = 1$, we recover a Poisson scatter. 

First, we model the expectation value $\langle N_{g}|\delta_{R, m}\rangle$ using two assumptions: an Eulerian quadratic bias model and the Lagrangian Gaussian model presented in~\cite{stucker_gaussian_2024, gould_cosmology_2024}. We focus on these two models as they give the closest fits to the simulations (compared to e.g. a simple linear bias model or a Lagrangian quadratic bias model). The Eulerian bias model is given by:
\begin{equation}
    \langle \delta_{R, g} | \delta_{R, m} \rangle = b_{1}^E \delta_{R, m} + \frac{b_{2}^E}{2} (\delta_{R, m}^2 - \sigma_{R, m}^2)
\end{equation}
while the Lagrangian Gaussian model is parametrized with:
\begin{equation}
    \langle \delta_{R, g} | \delta_{R, m} \rangle = \frac{(1 + \delta_{R, m})f_L\left(\delta_{L}(\delta_{R, m})\right)}{\langle (1 + \delta_{R, m})f_L\left(\delta_{L}(\delta_{R, m})\right) \rangle} - 1
\end{equation}
where $\delta_{L}(\delta_{R, m}) = \tau(\rho_{R, m} = 1 + \delta_{R, m})$ is given by spherical collapse (see equation~\eqref{eq:spherical_collapse}), and $f_L$ is defined by:
\begin{equation}
    f_L\left(\delta_{L}\right)=\frac{\exp \left[-\frac{\left(b_1^{G}\right)^2}{2 b_2^{G}}\right]}{\sqrt{1+b_2^{G} \sigma_{R, m}^2}} \exp \left[\frac{b_2^{G}\left(\frac{b_1^{G}}{b_2^{G}}+\delta_L\right)^2}{2\left(1+b_2^{G} \sigma_{R, m}^2\right)}\right].
\end{equation}

\begin{figure}
\begin{subfigure}{0.5\textwidth}
\includegraphics[scale=0.8]{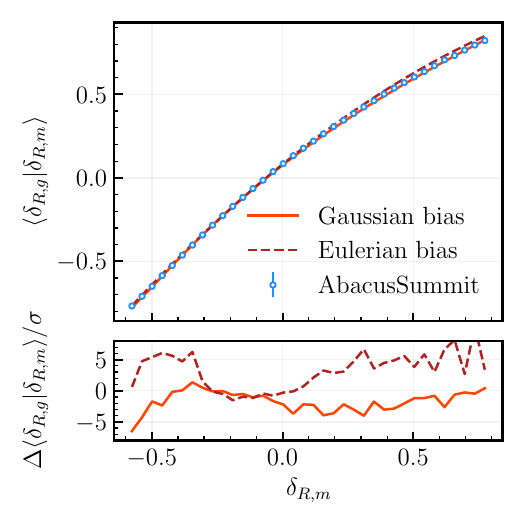}
\end{subfigure}
\begin{subfigure}{0.5\textwidth}
\includegraphics[scale=0.8]{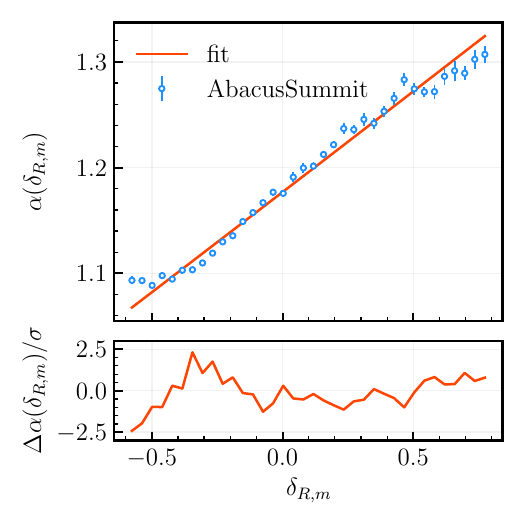}
\end{subfigure}
\caption{\textit{Left:} expectation value of $\delta_{R, g}$ conditioned on $\delta_{R,m}$, with the best-fit Eulerian quadratic and Lagrangian Gaussian models. $\delta_{R,m}$ is measured from 8 AbacusSummit simulations with all dark matter particles, and $\delta_{R,g}$ from the same simulations with halos populated with ELGs ($\nbar_g = 0.002 \; (\Mpch)^3$). $\delta_{R, m}$ values are binned by a factor 200 with respect to the original discrete values. Bottom panel shows the residuals between the models and the simulations divided by the standard deviation across the simulations. \textit{Right:} $\alpha_m(\delta_{R,m})$ measured from the 8 simulations and best-fit model from equation~\eqref{eq:alpha_shotnoise}. Bottom panel shows the residuals between the model and the measurement divided by the standard deviation across the simulations. The smoothing radius is $R = 10 \; \hMpc$ in both figures.}
\label{fig:galaxy_bias}
\end{figure}

The left panel of figure~\ref{fig:galaxy_bias} shows the conditional expectation value of $\delta_{R, g}$ given $\delta_{R, m}$, as measured from 8 simulations. We use 8 dark matter simulations with all the particles available ($\nbar \sim 1.24 \; (\Mpch)^3$) to compute $\delta_{R, m}$ with minimal sampling noise. We fix $\sigma_{R,m}$ to the value previously fitted on dark-matter only simulations ($\sigma_{R,m} = 0.48$). We fit the Eulerian quadratic bias model and the Lagrangian Gaussian bias model only in the range $-0.6 \leq \delta_{R, m} \leq 0.8 $, similarly to what was done in~\cite{gould_cosmology_2024}, as we find that both models fail to fit larger density ranges. Best-fit values for the bias parameters are reported in table~\ref{tab:bestfit_bias}.

Now let us model the scatter around this expectation value. As in~\cite{gould_cosmology_2024}, we assume a quadratic polynomial form for $\alpha(\delta_{R, m})$:
\begin{equation}
    \alpha(\delta_{R, m}) = \alpha_0 + \alpha_1 \delta_{R, m} + \alpha_2 \delta_{R, m}^2.
\label{eq:alpha_shotnoise}
\end{equation}
The right panel of figure~\ref{fig:galaxy_bias} shows the measurement of $\alpha(\delta_{R, m}) = \nbar_g V_R \langle \delta_{R, g}^2 | \delta_m \rangle / (1 + \langle \delta_{R, g} | \delta_m \rangle)$ 
from the 8 simulations compared to the best-fit model~\eqref{eq:alpha_shotnoise}. Best-fit values for $\alpha_0$, $\alpha_1$ and $\alpha_2$ are reported in table~\ref{tab:bestfit_bias}. We find that $\alpha(\delta_{R, m})$ deviates strongly from Poissonian shot noise, as was already noted in~\cite{friedrich_pdf_2022, gould_cosmology_2024}.

\begin{table}
\centering
\begin{tabular}{|c|c||c|c||c|c|c|}
\hline
            \multicolumn{2}{|c||}{Eulerian quadratic bias}   & \multicolumn{2}{|c||}{Lagrangian Gaussian bias} & \multicolumn{3}{|c|}{Shot noise} \\
\hline
            $b_1^E$   & $b_2^E$   & $b_1^G$   & $b_2^G$ & $\alpha_0$  &  $\alpha_1$  & $\alpha_2$\\
\hline
            1.27      & -0.71    & 0.10      & -0.74    & 1.1  & 0.19  & 0.001 \\
\hline
\end{tabular}
\caption{Best-fit values for the parameters of the quadratic Eulerian bias model, the Gaussian Lagrangian model and the shot noise model $\alpha(\delta_{R,m})$.}
\label{tab:bestfit_bias}
\end{table}

\begin{figure}
\begin{subfigure}{0.5\textwidth}
\includegraphics[scale=0.8]{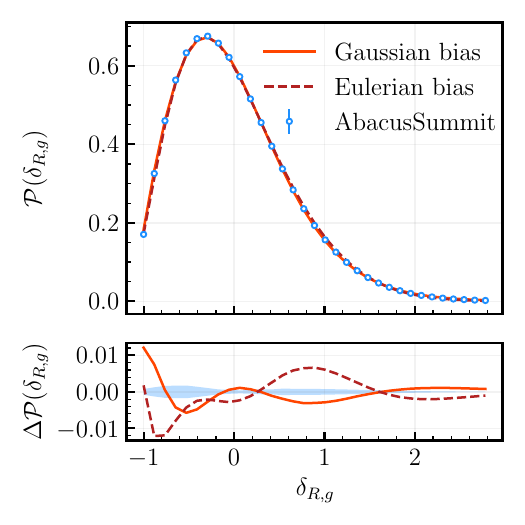}
\end{subfigure}
\begin{subfigure}{0.5\textwidth}
\includegraphics[scale=0.8]{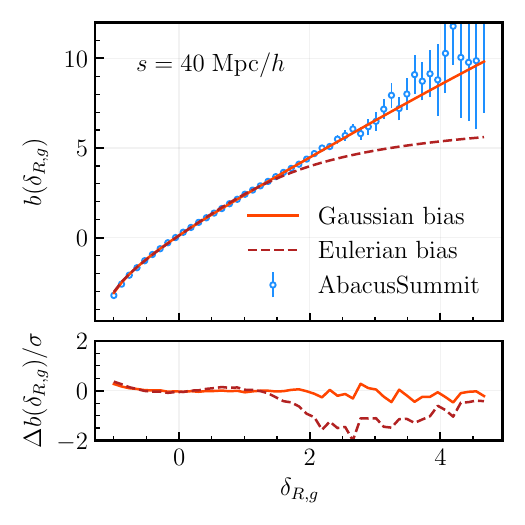}
\end{subfigure}
\caption{\textit{Left:} average one-point PDF of the galaxy count-in-cell density measured on 25 AbacusSummit simulations populated with ELGs ($\nbar_g = 0.002 \; (\Mpch)^3$), compared to the LDT prediction for the matter PDF and best-fit Gaussian (solid line) and Eulerian (dashed line) bias models. Bottom panel shows the difference between the model and the mocks. \textit{Right:} average bias function of the galaxy count-in-cell density at separation $s = 40 \; \hMpc$, compared to the LDT prediction and best-fit Gaussian and Eulerian bias models. Bottom panel shows the difference between the model and the mocks divided by the standard deviation of the 25 mocks. The smoothing radius is $R = 10 \; \hMpc$ in both figures.}
\label{fig:one-point_pdf_ELG}
\end{figure}

In the left panel of figure~\ref{fig:one-point_pdf_ELG}, we show the comparison between the measured count-in-cell PDF of the 25 ELG simulations and the LDT model convolved with Eulerian and Gaussian bias models (with bias and shot noise parameters from table~\ref{tab:bestfit_bias}). The model fits relatively well the simulations, although not at the level of their cosmic variance (blue area).

The two-point PDF for a biased tracer is now written:
\begin{equation}
    \prob_g(N_g, N_g^{\prime}) 
    = \int d\delta_{R, m} \int d\delta_{R, m}^{\prime} \prob(N_g |\delta_{R, m}) \prob(N_g^{\prime}|\delta_{R, m}^{\prime}) \prob_m(\delta_{R, m}, \delta_{R, m}^{\prime}),
\end{equation}
where the LDT prediction for $\prob(\delta_{R, m}, \delta_{R, m}^{\prime})$ is given by equation~\eqref{eq:bias_peak_background}, such that:
\begin{equation}
    \prob_g(N_g, N_g^{\prime}) 
    = \prob_g(N_g)\prob_g(N_g^{\prime}) 
    \left[
    1 + \xi_{R, m}(\vs) \tilde{b}_{g}(\delta_{R, g}) \tilde{b}_{g}(\delta_{R, g}^{\prime})
    \right],
\label{eq:peak_background_galaxies}
\end{equation}
where:
\begin{equation}
    \tilde{b}_{g}(\delta_{R, g}) = \nbar_g V_R \frac{\int d\delta_{R, m} \prob(N_g|\delta_{R, m})\prob_m(\delta_{R, m})b_m(\delta_{R, m})}{\prob_g(\delta_{R, g})}.
\end{equation}
Rewriting the bias function definition from equation~\eqref{eq:bias_peak_background} for biased tracers:
\begin{equation}
    \prob_g(\delta_{R, g}, \delta_{R, g}^{\prime}) 
    = \prob_g(\delta_{R, g})\prob_g(\delta_{R, g}^{\prime}) 
    \left[
    1 + \xi_{R, g}(\vs) {b}_{g}(\delta_{R, g}, \vs) {b}_{g}(\delta_{R, g}^{\prime}, \vs)
    \right],
\end{equation}
we define the galaxy bias function $b_{g}$ by~\cite{uhlemann_question_2018}:
\begin{equation}
    b_{g}(\delta_{R, g}, \vs) = \tilde{b}_{g}(\delta_{R, g}) \sqrt\frac{\xi_{R,m}(\vs)}{\xi_{R,g}(\vs)}.
\label{eq:bias_function_shotnoise_galaxies}
\end{equation}
The right panel of figure~\ref{fig:one-point_pdf_ELG} shows this bias function model (with bias and shot noise parameters from table~\ref{tab:bestfit_bias}) compared to the ELG simulations, at separation $s = 40 \; \hMpc$. Here we estimate $\xi_{R,m}(\vs)$ and $\xi_{R,g}(\vs)$ from the simulations, from estimator~\eqref{eq:smoothed_corr_estimator}. The Gaussian bias model is in very good agreement with the simulations, within their cosmic variance, while the Eulerian bias model slighlty underestimates the bias function for large densities.

For density-splits, we want the only input from simulations to be the matter smoothed two-point correlation function, so use the following relation, which comes from integrating equation~\eqref{eq:peak_background_galaxies}:
\begin{equation}
    \xiRgDS(\vs) = \frac{\int_{\DS} d\delta_{R, g} \tilde{b}_g(\delta_{R, g}) \prob_g(\delta_{R, g})}
    {\int_{\DS} d\delta_{R, g} \prob_g(\delta_{R, g})} 
    \langle \delta_{R, g}  \tilde{b}_g(\delta_{R, g}) \rangle \xi_{R, m}(\vs)
\end{equation}
where $\langle \delta_{R, g}  \tilde{b}_g(\delta_{R, g}) \rangle =\int d\delta_{R, g} \delta_{R, g}  \tilde{b}_g(\delta_{R, g}) \prob(\delta_{R, g})$.
Figure~\ref{fig:density_splits_ELG} shows the density-split measurements (with three density-splits) from the ELG simulations, together with the LDT predictions with the Eulerian and Gaussian bias models. The prediction now includes the matter two-point correlation function, which here we measure from the dark matter simulations with $\nbar = 0.0034 \; (\Mpch)^3$ using Landy-Szalay estimator. The Gaussian bias model yields much better agreement with the simulations than the Eulerian bias model. With the Gaussian bias model, the biased LDT density-split prediction is within the cosmic variance of the simulations at scales {$s \gtrsim 40 \; \hMpc$.

\begin{figure}
\begin{subfigure}{0.5\textwidth}
\includegraphics[scale=0.8]{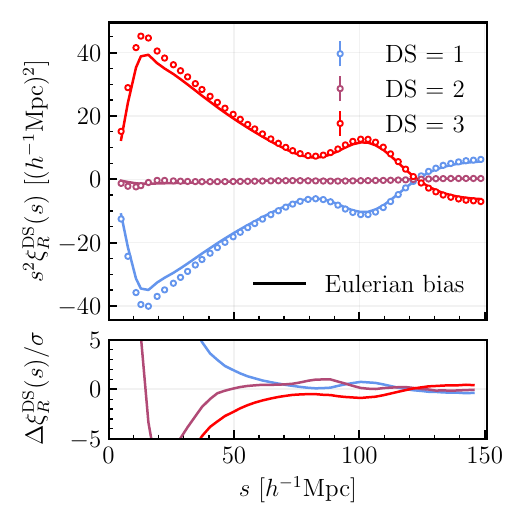}
\end{subfigure}
\begin{subfigure}{0.5\textwidth}
\includegraphics[scale=0.8]{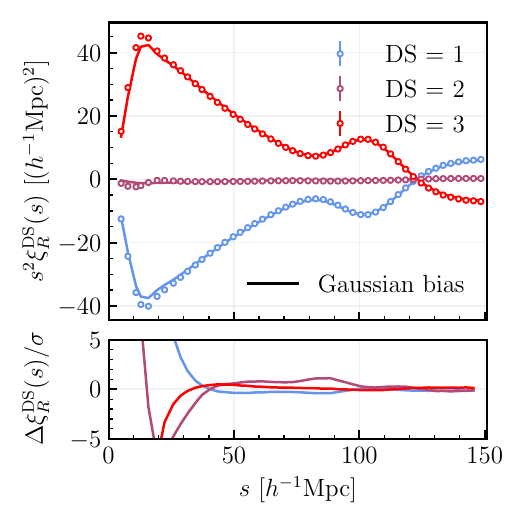}
\end{subfigure}
\caption{Average density-split correlation functions measured from the 25 AbacusSummit simulations populated with ELGs ($\nbar = 0.002 \; (\Mpch)^3$). Solid lines show the LDT prediction with quadratic Eulerian (left) and Gaussian Lagrangian (right) bias models. Bottom panels show the residuals between the model and simulations, divided by the standard deviation across the 25 simulations. The smoothing radius is $R = 10 \; \hMpc$ in both figures.}
\label{fig:density_splits_ELG}
\end{figure}

\section{Conclusions}\label{sec:conclusion}

We have introduced a formalism to predict density-split correlation functions -- in particular, the cross-correlation of density-splits with the matter field -- with top-hat smoothing in real space, for dark matter. In the general case, we have shown that density-split correlation functions can be factorized in two components: a bias factor, a priori scale-dependent, which multiplies the smoothed two-point correlation function. This means that all the additional information to the standard two-point correlation function contained in density-splits comes from this bias factor, which is a combination of 
the one-point PDF of the density and its bias function at the given separation.

We first presented a derivation of this bias factor in the Gaussian case, where it is independent of scale and only a function of the one-point PDF: the average smoothed density contrast in the density-split region $\DS$ divided by the variance of the smoothed density contrast. To validate our predictions, we computed density-split correlation functions, for three or five density-splits, on 25 AbacusSummit N-body simulations at redshift $z = 0.8$. We showed that the Gaussian prescription is not sufficient to describe these realistic simulations. 

Additionally, we introduced the bias function and density-split predictions in the case of a joint shifted log-normal distribution for the two-point density. We also accounted for shot noise by convolving this two-point log-normal PDF with two independent Poisson distributions. Although the one-point PDF of the simulations is very well described by the log-normal assumption and Poisson shot noise -- provided that the variance and skewness of the log-normal distribution are adjusted accordingly, the log-normal model tends to over-predict the bias function in the region $\delta_R \simeq 0$, and under-predict in for larger absolute values of $\delta_R$. As a result, for the density-split correlation functions, the log-normal model is not at the level of the statistical error of the simulations. 

Finally, we have built upon previous works modeling the PDF of count-in-cell density and its bias function in the large separation limit with LDT~\cite{codis_large-scale_2016}, to derive a model for density-split correlation functions. As for the log-normal model, we added a Poisson shot noise to the LDT predictions. We find that, when accounting for shot noise, the LDT prediction for the bias function is in excellent agreement with the mocks, as long as we consider sufficiently large separations compared to the smoothing radius (in our case, $s \gtrsim 40 \; \hMpc$ for a smoothing radius of $R = 10 \; \hMpc$). Therefore, the LDT model for density-splits performs much better than the log-normal model at separations $s \gtrsim 50 \; \hMpc$. The mismatch at lower separations is expected as the bias function prediction is only valid for large separations before the smoothing radius. Note however that the agreement remains similar for $s \gtrsim 50 \; \hMpc$ even with a smoothing radius of $25 \; \hMpc$. We compared our model predictions against simulations with two different average densities: a high number density ($\nbar = 0.0034 \; (\Mpch)^3$), and a lower number density ($\nbar = 5 \times 10^{-4} \; (\Mpch)^3$), which is more typical of a DESI DR1 sample. The agreement is good at separations $s \gtrsim 50 \; \hMpc$ for both densities.

Notably, the bias factor in the density-split predictions does not depend on the separation, in the large separation limit, i.e. here for $s \gtrsim 50 \; \hMpc$. At lower separations, the bias depends on scale and its behavior is not captured by LDT.

In addition, we have extended our formalism to the case of biased tracers with a Gaussian Lagrangian bias, which we compared to a quadratic Eulerian bias, and non-Poisson shot noise, in line with previous studies~\cite{uhlemann_question_2018, stucker_gaussian_2024, gould_cosmology_2024}. We tested our prediction against dark matter simulations populated with ELGs following a HOD prescription. The extended model with Gaussian Lagrangian bias and non-Poisson shot noise shows good agreement with the simulations on large scales, similarly to the dark matter-only case.

Using the LDT framework, we developed a reliable model of density-splits at large separations in real space, which is key to understand where the gain in sensitivity of the density-splits comes from. Nevertheless, further work is still needed to turn it into an analysis tool for density-split clustering, which requires to operate in redshift space.

Another possible avenue of research is to extend the LDT framework to include physics beyond $\Lambda$CDM, and quantify the impact of the neutrino mass~\cite{uhlemann_fisher_2020}, modified gravity~\cite{cataneo_matter_2022} or primordial non-Gaussianities~\cite{friedrich_primordial_2020} on density-split statistics.

\appendix

\section{Effect of shot noise}\label{sec:shotnoise}

In this section we present the effect of the convolution with a Poisson shot noise of the one-point PDF and bias function. Figure~\ref{fig:shotnoise} shows the LDT predictions for the one-point PDF and bias function before and after the convolution with a Poisson distribution. We see that shot noise has a big effect on the one-point PDF, even for the higher density $\bar{n} = 0.0034 \; (\Mpch)^3$. At the level of the bias function, the shot noise flattens the bias function, with higher bias values for negative density contrast, and lower bias values for large density contrast.

\begin{figure}
\begin{subfigure}{0.5\textwidth}
\includegraphics[scale=0.8]{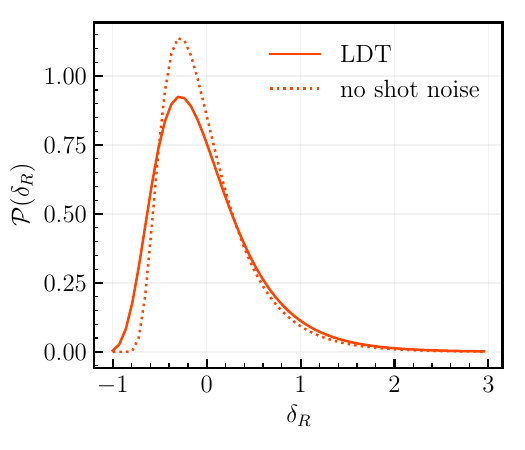}
\end{subfigure}
\begin{subfigure}{0.5\textwidth}
\includegraphics[scale=0.8]{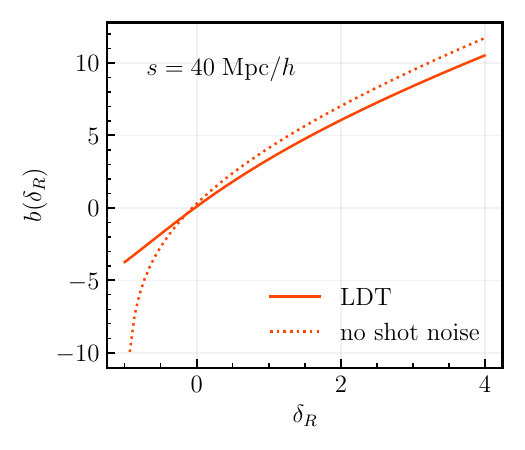}
\end{subfigure}
\caption{Comparison between the LDT model before (dotted line) and after (solid line) the convolution with a Poisson distribution. The variance of the LDT model is fitted to the simulations with $\bar{n} = 0.0034 \; (\Mpch)^3$ and $R = 10 \; \hMpc$.}
\label{fig:shotnoise}
\end{figure}

\acknowledgments

We thank Alexandre Barthelemy for insightful discussions, and Antoine Rocher for providing us with the code to generate the ELG HOD model.

\bibliographystyle{JHEP}
\bibliography{references}{}

\end{document}